\documentclass[iop]{emulateapj}
\usepackage{graphicx}
\usepackage{epsfig}
\usepackage{epstopdf}
\usepackage{natbib}
\usepackage{subfigure}
\usepackage{amsmath} 
\usepackage{array}
\usepackage{rotating}

\graphicspath{{./figures/}} 

\begin{document}

%
%
  %
  %
\title{A Fast and Accurate Method to Capture the 
Solar Corona/Transition Region Enthalpy Exchange}
\shorttitle{A Fast and Accurate Method to Capture the 
Corona/Transition Region Enthalpy Exchange}
\author{
  C. D. Johnston\altaffilmark{1}   \& 
  S. J. Bradshaw\altaffilmark{2}
  }
\shortauthors{Johnston \& Bradshaw}
\affil{\altaffilmark{1}School of Mathematics and Statistics, 
  University of St Andrews, St Andrews, Fife, KY16 9SS, UK;
  cdj3@st-andrews.ac.uk
  }
\affil{\altaffilmark{2}Department of Physics and Astronomy, Rice University, Houston, TX 77005, USA; stephen.bradshaw@rice.edu 
}
  %
  %
%
%
  %
  %
\begin{abstract}
  The brightness of the emission from coronal loops in the
  solar atmosphere is strongly dependent on the temperature and 
  density of the confined plasma. 
  After a release of energy, these loops undergo a heating and 
  upflow phase, followed by a cooling and downflow cycle.
  Throughout, there are significant variations in the 
  properties of the coronal plasma. In particular, the 
  increased coronal 
  temperature leads to an excess downward heat flux that the 
  transition region is unable to radiate. This generates an 
  enthalpy flux from the transition region to the corona,
  increasing the coronal density. 
  The enthalpy exchange is highly sensitive to the transition 
  region 
  resolution in numerical simulations. With a numerically 
  under-resolved transition region, major errors occur in 
  simulating
  the coronal density evolution and, thus, the 
  predicted loop emission. This Letter presents a new method 
  that addresses the difficulty of obtaining the correct 
  interaction between the corona and corona/chromosphere 
  interface. In the transition region, an adaptive thermal 
  conduction approach is used that broadens any unresolved 
  parts of the atmosphere. We show that this approach, referred 
  to as TRAC, successfully removes the influence of numerical 
  resolution on the coronal density response to heating while 
  maintaining high levels of agreement with fully resolved 
  models. 
  When employed with coarse spatial 
  resolutions, typically achieved in multi-dimensional MHD 
  codes, the peak density errors are less than 3\% and the 
  computation time
  is three orders of magnitude faster than fully resolved 
  field-aligned models.
  The advantages of using TRAC in field-aligned hydrodynamic 
  and multi-dimensional magnetohydrodynamic simulations are 
  discussed. 
\end{abstract}
  %
  %
%
%
  %
  %
\section{Introduction}
  \label{Sect:intro}
  \indent
  One of the main difficulties encountered when using
  field-aligned hydrodynamic models to study the 
  physics of magnetically closed coronal loops 
  is the need to implement a grid that 
  fully resolves the steep and dynamic transition region (TR) 
  \citep[see e.g.][hereafter BC13]{paper:Bradshaw&Cargill2013}.
  In a static coronal loop, the energy balance in the TR
  is between a downward heat flux from the corona and
  optically thin radiative losses to space.   
  Defining the temperature length scale as 
  $L_T =T/|dT/ds|$,
  where $s$ is a coordinate along the magnetic 
  field, such  static  
  loops can have $L_T$ of order 1 km
  in the TR. During impulsive heating, the heat flux from
  the corona can be enhanced by orders of magnitude, so
  that $L_T$ becomes very small, of order 100~m
  and less in some cases.
  \\
  \indent
  Properly
  resolving these small length scales
  is essential in order to 
  obtain the 
  correct coronal density in response to heating (BC13),
  otherwise the downward
  heat flux 
  \lq jumps\rq\ over an under-resolved TR to the 
  chromosphere, where the incoming
  energy is then strongly radiated.
  BC13 showed 
  that major errors in simulating the coronal density 
  evolution were likely with 
  a lack of spatial resolution.
  \\
  \indent  
  However, since the thermal conduction time scale across
  a grid cell scales as $(\Delta s)^2$ and 
  $T^{-5/2}$ (where $\Delta s$ is the grid cell width), 
  cells in the cooler lower TR must become
  dramatically smaller to resolve the temperature
  gradient that supports the incoming heat flux. 
  In addition, numerical stability requires that
  the minimum time scale across the entire grid is 
  temporally resolved. Thus,
  locally satisfying $\Delta s < L_T$ in the TR
  implies 
  long computation times for fully resolved 
  field-aligned simulations.
  The problem is more severe in
  three-dimensional~(3D)
  magnetohydrodynamic (MHD)
  models where computational resources place significant
  constraints on the achievable resolutions.
  Therefore, other approaches need to be found.
  \\
  \indent
  One such approach has been developed by 
  \cite{paper:Linkeretal2001},
  \cite{paper:Lionelloetal2009} and 
  \cite{paper:Mikicetal2013} for use in global 3D MHD
  simulations. 
  This technique 
  modifies the temperature dependence of the parallel
  thermal conductivity $\kappa_\parallel(T)$ 
  and radiative emissivity $\Lambda(T)$ 
  below a fixed temperature value ($T_c$) in the TR.
  It has been demonstrated that such modifications 
  artificially broaden the TR for $T<T_c$ without 
  significantly changing the coronal properties of the loop.
  While this is a particularly attractive property the method 
  still requires sufficiently high resolution
  to properly resolve the small length scales in the TR above
  $T_c$.
  Furthermore, 
  the characteristic temperature of the TR
  can change dramatically during the heating and cooling
  cycle, making it unclear how suitable a fixed value of
  $T_c$
  is for capturing the  
  dynamic evolution of coronal loops that are 
  impulsively heated.
  \\
  \indent
  Another approach has been developed by
  \cite{paper:Johnstonetal2017a,paper:Johnstonetal2017b},
  who treated the (unresolved) TR as a discontinuity 
  across which energy is conserved through the
  imposition of a jump condition.
  No attempt is made to properly resolve the TR.
  Instead, the method is employed with coarse spatial
  resolutions that are supported
  by dynamically locating the top 
  of the unresolved transition region (UTR).
  This approach gives
  good agreement with 
  fully resolved HYDRAD 
  simulations
  \citep[][BC13]{paper:Bradshaw&Mason2003,
  paper:Bradshaw&Cargill2006} 
  whilst significantly 
  speeding up the computation time
  (e.g.~Johnston et al., submitted).
  \\
  \indent
  This Letter proposes a new approach for modelling the 
  TR that incorporates the strengths of both approaches. 
  Principally, the ability to dynamically identify
  the UTR to prescribe an adaptive $T_c$ and then broaden the
  length scales in this unresolved region only
  in order to eliminate the need for 
  highly resolved numerical grids.
  The outcome is an extremely powerful method
  that (1) removes the influence of numerical resolution
  on the coronal response to heating
  and (2) accurately predicts
  the results of properly resolved 
  field-aligned models (e.g.~BC13)
  when employed with the coarser spatial resolutions
  achieved by multi-dimensional 
  MHD codes.
  %
  %
%
%
  %
  %
\section{Numerical Model and Experiments
  \label{Sect:model}}
  %
  %
%
%
  %
  %
\subsection{Numerical Model
  \label{Sect:num_model}}
  \indent
  We solve the field-aligned 
  hydrodynamic equations using the
  HYDRAD code  
  \citep[][BC13]{paper:Bradshaw&Mason2003,
  paper:Bradshaw&Cargill2006} run in single fluid mode.
  HYDRAD uses adaptive regridding to ensure adequate spatial 
  resolution, with the grid being 
  refined such that cell-to-cell 
  changes in the temperature and density are kept between 
  user defined values 
  (taken as 5\% and 10\% here), 
  where possible. The largest grid cell in all of
  our calculations has a width of $10^6$ m (1,000 km)
  and each successive refinement splits the cell into two.
  Thus, a refinement level of RL leads to cell widths decreased 
  by $1/2^{\textrm{RL}}$. 
  The maximum value of RL is limited to 14, corresponding 
  to a grid cell width of 61 m
  in the most highly resolved parts of the TR. 
  %
  %
%
%
  %
  %
\subsection{Transition Region Adaptive Conduction
  \label{Sect:TRAC}}
  \indent
  While it is possible to employ such high resolution grids 
  to properly resolve
  the TR in field-aligned codes,  this approach
  is unlikely to be a viable way of
  running multi-dimensional MHD simulations.
  Therefore, we have developed a new method  
  that addresses this difficulty of obtaining adequate spatial
  resolution in numerical simulations
  by modelling the 
  transition region using adaptive conduction (TRAC) 
  coefficients
  that act to
  broaden any unresolved parts of the atmosphere.
  This treatment of thermal conduction in the TR,
  referred to as the TRAC method,
  relies on a
  dynamic capability to track resolved and modify unresolved
  conductive fluxes.
  %
  %
%
%
  %
  %
\subsubsection{Identification of an adaptive cutoff temperature
  \label{Sect:TRAC_identification}}
  \indent
  The implementation of TRAC is comprised of two main parts.
  The first part is to identify the maximum  
  temperature of any unresolved grid cells
  in the numerical domain and use the
  calculated value to prescribe an adaptive cutoff 
  temperature ($T_c$).
  This is done by using an algorithm that is 
  based on the method employed by 
  \cite{paper:Johnstonetal2017a,paper:Johnstonetal2017b}
  for locating the top of the 
  UTR in the
  jump condition method as follows.
  \\
  \indent
  We define the temperature length scale as,
  \begin{align}
    L_T(T(s)) = \frac{T}{|dT/ds|},
    \label{Eqn:L_T} 
  \end{align}
  and the resolution in a simulation is given by 
  the local grid cell width,
  \begin{align}
    L_R(s) = \Delta s,
    \label{Eqn:L_R}
  \end{align}
  where $s$ is the spatial coordinate along the magnetic field.  
  \\
  \indent
  Using these definitions, 
  the cutoff temperature is defined as 
  the maximum temperature that violates the resolution 
  criteria of
  \cite{paper:Johnstonetal2017a,paper:Johnstonetal2017b},
  \begin{align}
    T_c=\textrm{max}(T(s)) \,\Big|\,
    \frac{L_R(s)}{L_T(s)} > \delta=\frac{1}{2},
    \label{Eqn:T_c} 
  \end{align}
  which corresponds to not having multiple grid cells
  across the temperature length scale (i.e.~unresolved
  temperature gradients).
  \\
  \indent
  An upper bound for the cutoff temperature is set as 
  20\% of the peak 
  coronal temperature in the loop
  at the time when $T_c$ is evaluated
  (though the results are only weakly dependent on the 
  maximum temperature fraction),  
  \begin{align}
    T_{\textrm{max}}=0.2 \, T_{\textrm{peak}},
    \label{Eqn:T_c_max} 
  \end{align}
  and a lower bound set as the 
  temperature value of the isothermal
  chromosphere,
  \begin{align}
    T_{\textrm{min}}= T_{\textrm{chrom}}.
    \label{Eqn:T_c_min} 
  \end{align}
  In this Letter the lower bound is
  taken as
  $T_{\textrm{chrom}}=2\times 10^4$~K.
  Employing these definitions, we  dynamically 
  adjust $T_c$ with the criteria that it should satisfy,
  \begin{align}
    T_{\textrm{min}} \leq T_c \leq  T_{\textrm{max}}.
    \label{Eqn:T_c_criteria} 
  \end{align}
  Hence, the cutoff temperature is 
  adaptive in identifying the UTR,
  with the value of $T_c$ used in the method changing
  in response to coronal heating and cooling.
  %
  %
%
%
  %
  %
\subsubsection{Broadening the unresolved transition region
  \label{Sect:TRAC_broadening}}
  \indent
  The second part of TRAC 
  is to broaden the steep temperature and   
  density gradients in the UTR.
  This is achieved using the approach developed by 
  \cite{paper:Linkeretal2001},
  \cite{paper:Lionelloetal2009} and 
  \cite{paper:Mikicetal2013}
  \\
  \indent
  Below the
  adaptive cutoff temperature ($T_c$),
  the parallel thermal conductivity ($\kappa_\parallel$)
  is set to a constant value,
  \begin{align}
    \kappa_\parallel(T) &= \kappa_0 T^{5/2},
    \quad \forall T \geq T_c;
    \\
    \kappa_\parallel(T) &= \kappa_\parallel(T_c),
    \quad \forall T < T_c,
  \end{align}
  and the radiative loss rate ($\Lambda$)  
  is modified to preserve
  $\Lambda(T) \kappa_\parallel(T)$,
  \begin{align}
    \Lambda(T) &= \Lambda(T),
    \quad \forall T \geq T_c;
    \\
    \Lambda(T) &= \Lambda(T) 
    \left(\dfrac{T}{T_c}
    \right)^{5/2},
    \quad \forall T < T_c.
  \end{align}
  \indent
  Increasing the parallel thermal conductivity
  and decreasing the radiative loss
  rate, at temperatures below $T_c$, has the
  desired effect of
  broadening the temperature
  length scales in the UTR.
  This helps TRAC prevent the
  heat flux jumping across the unresolved region while 
  maintaining accuracy in the
  properly resolved parts of the atmosphere. 
  (The broadening effect will be explained in greater detail in 
  forthcoming work.)
  Furthermore, we note that the formulation of TRAC 
  (1) makes no assumptions about the spatial
  resolution in a simulation and 
  (2) the parallel thermal conductivity
  reduces to the classical Spitzer-Harm value 
  when the TR is properly resolved.
  %
  %
  %
  %
\begin{sidewaysfigure*}
  \vspace{0.4\linewidth}
  \hspace*{-0.075\linewidth}
  \subfigure{\includegraphics[width=0.36\linewidth]
  {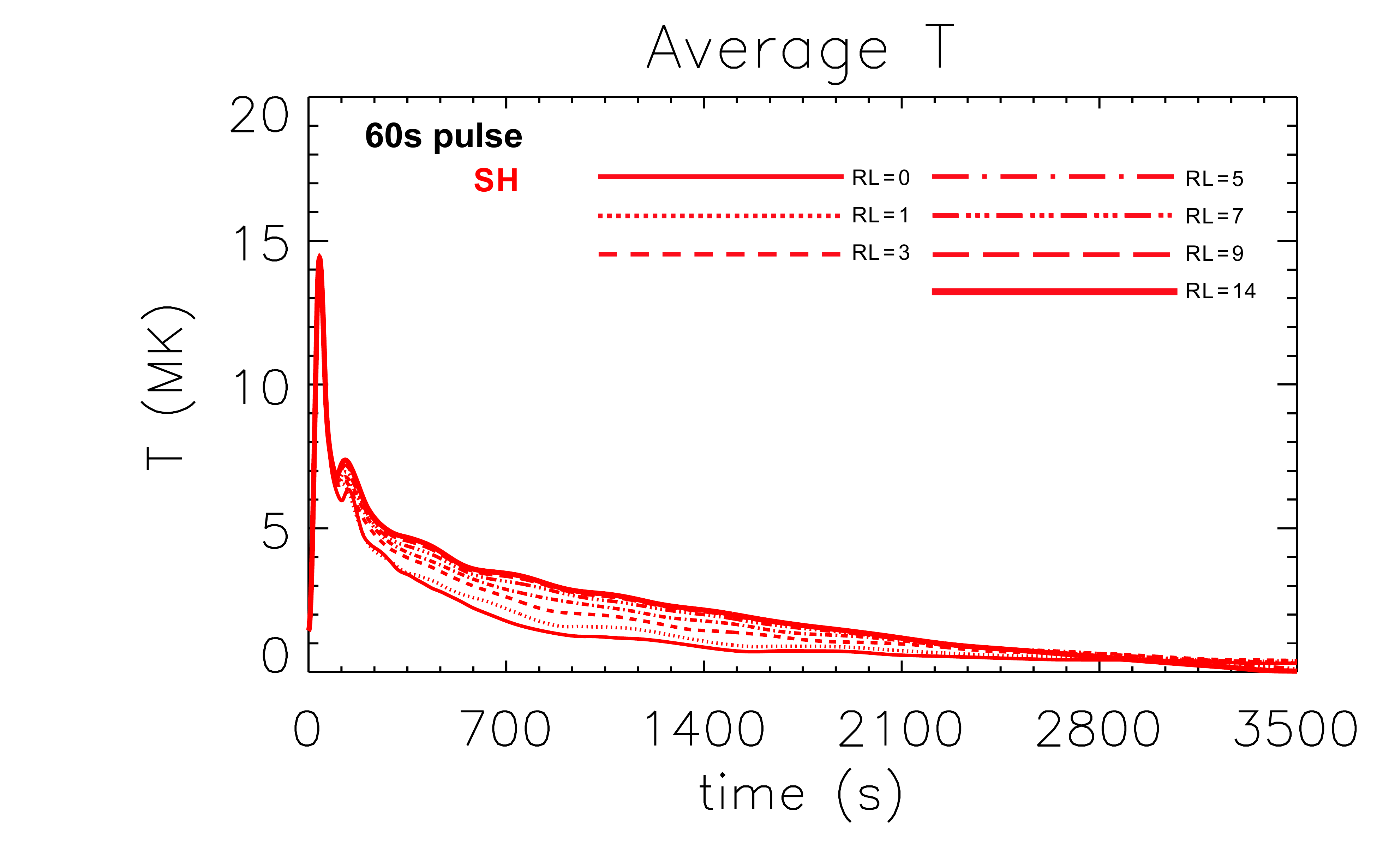}}
  \hspace*{-0.005\linewidth}
  \subfigure{\includegraphics[width=0.36\linewidth]
  {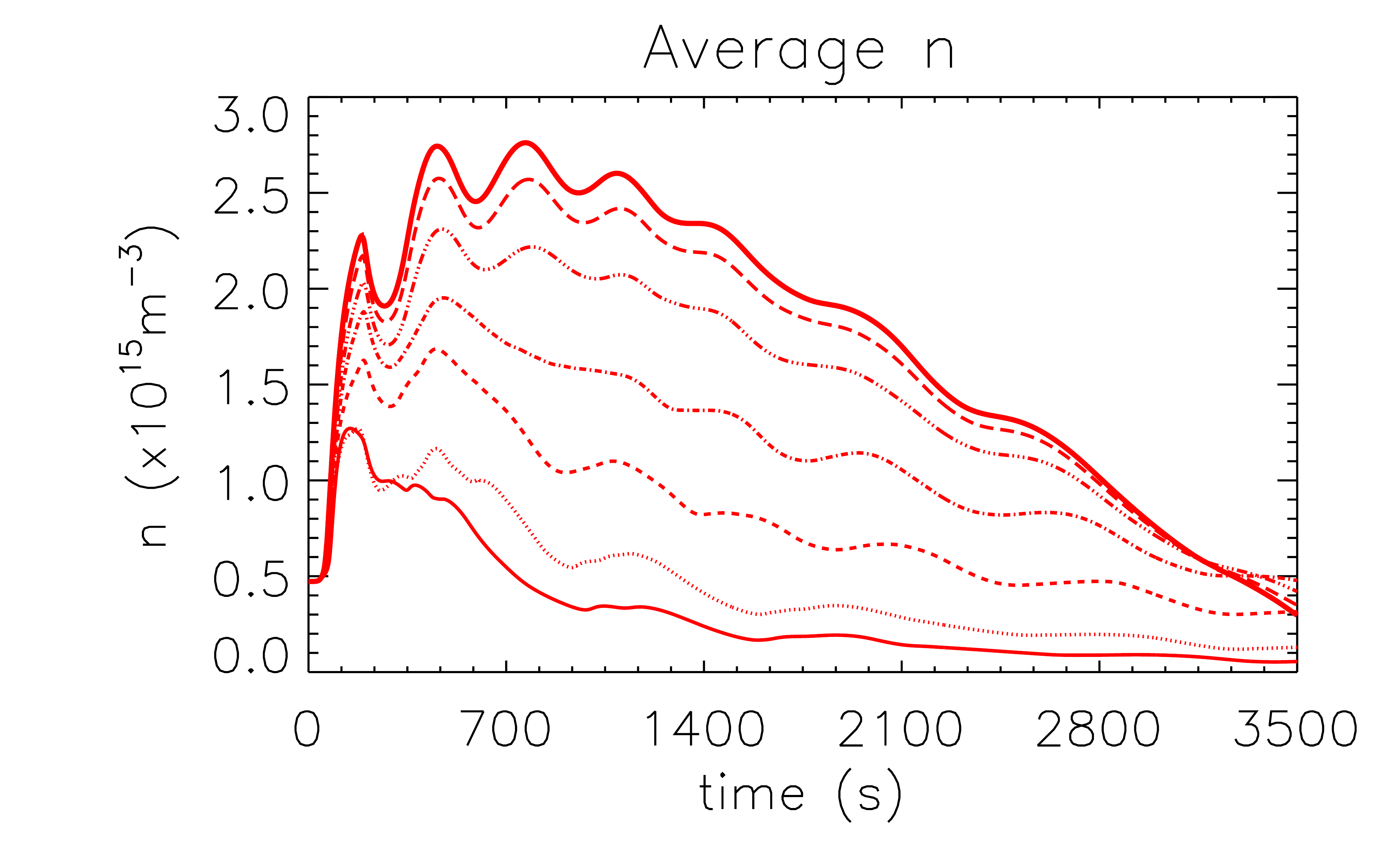}}
  \hspace*{-0.005\linewidth}
  \subfigure{\includegraphics[width=0.36\linewidth]
  {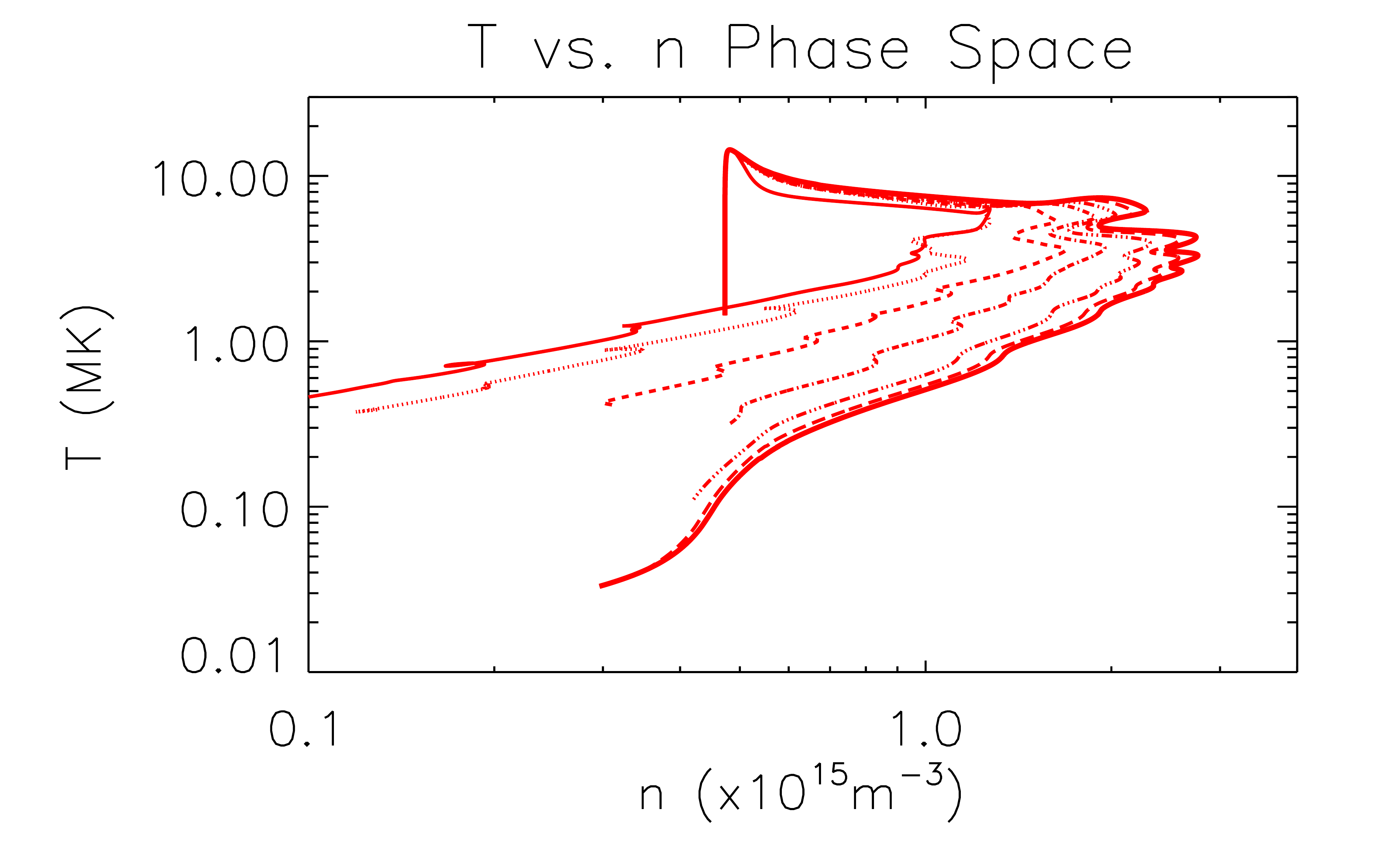}}
  \\
  \hspace*{-0.075\linewidth}
  \subfigure{\includegraphics[width=0.36\linewidth]
  {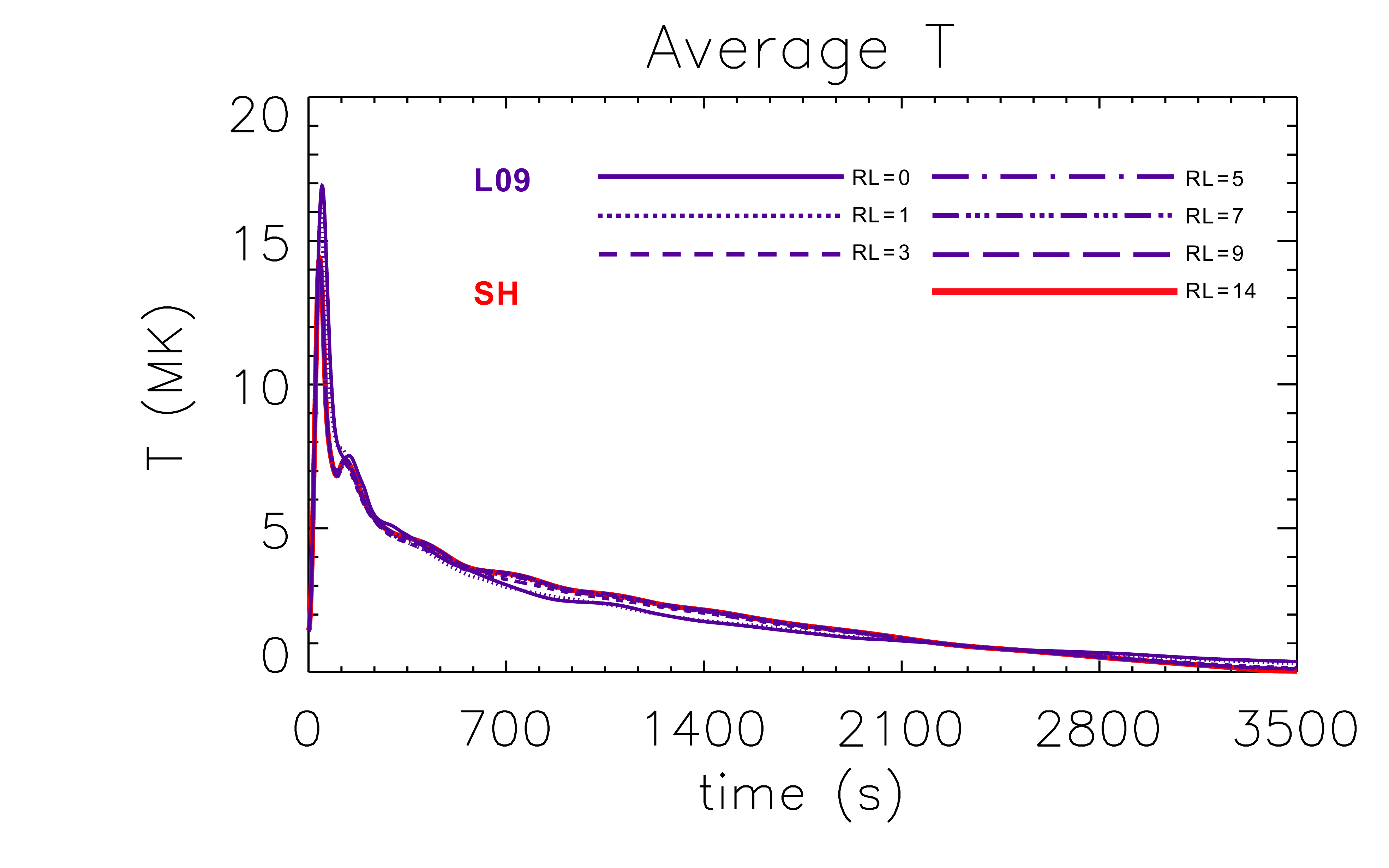}}
  \hspace*{-0.005\linewidth}
  \subfigure{\includegraphics[width=0.36\linewidth]
  {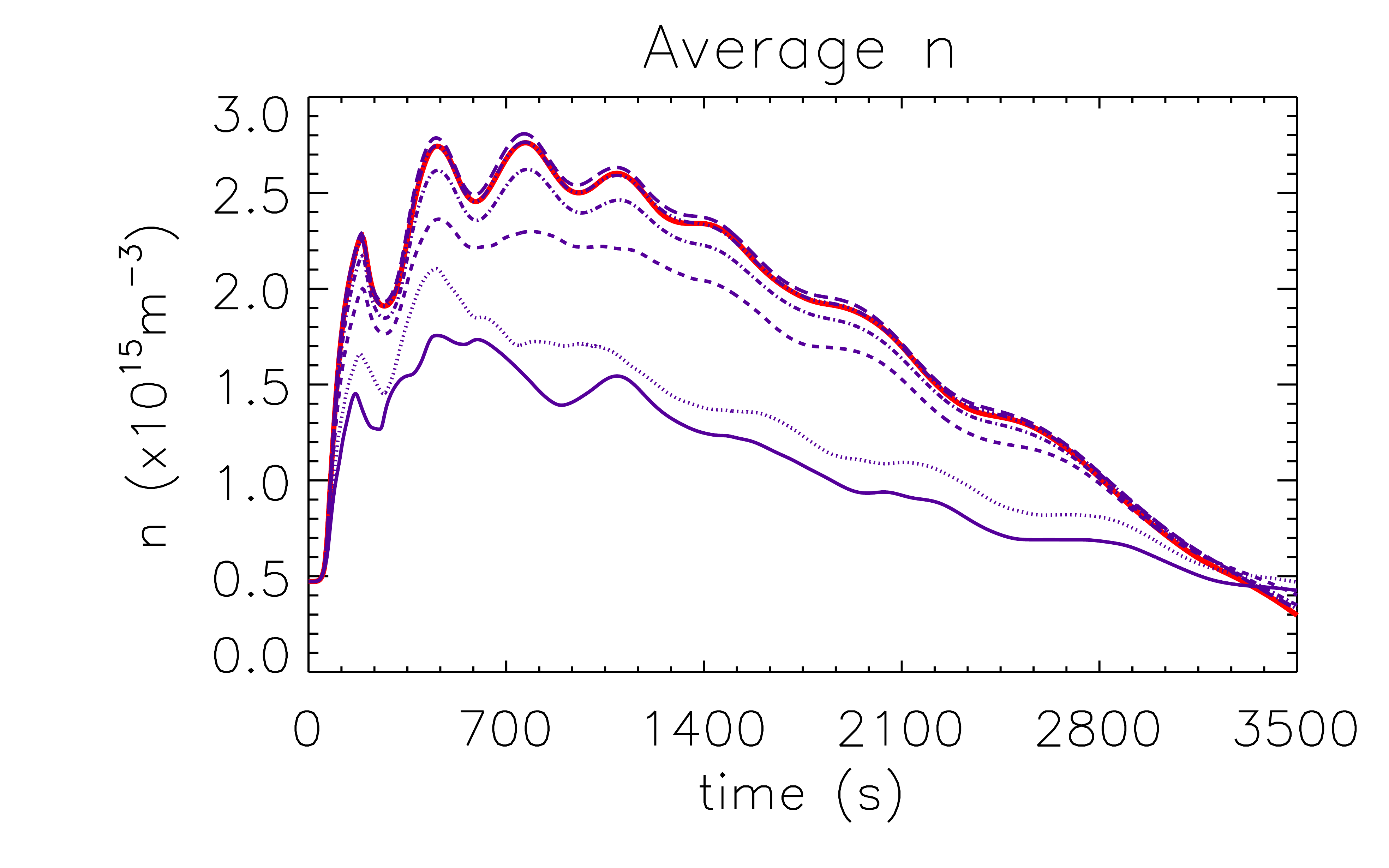}}
  \hspace*{-0.005\linewidth}
  \subfigure{\includegraphics[width=0.36\linewidth]
  {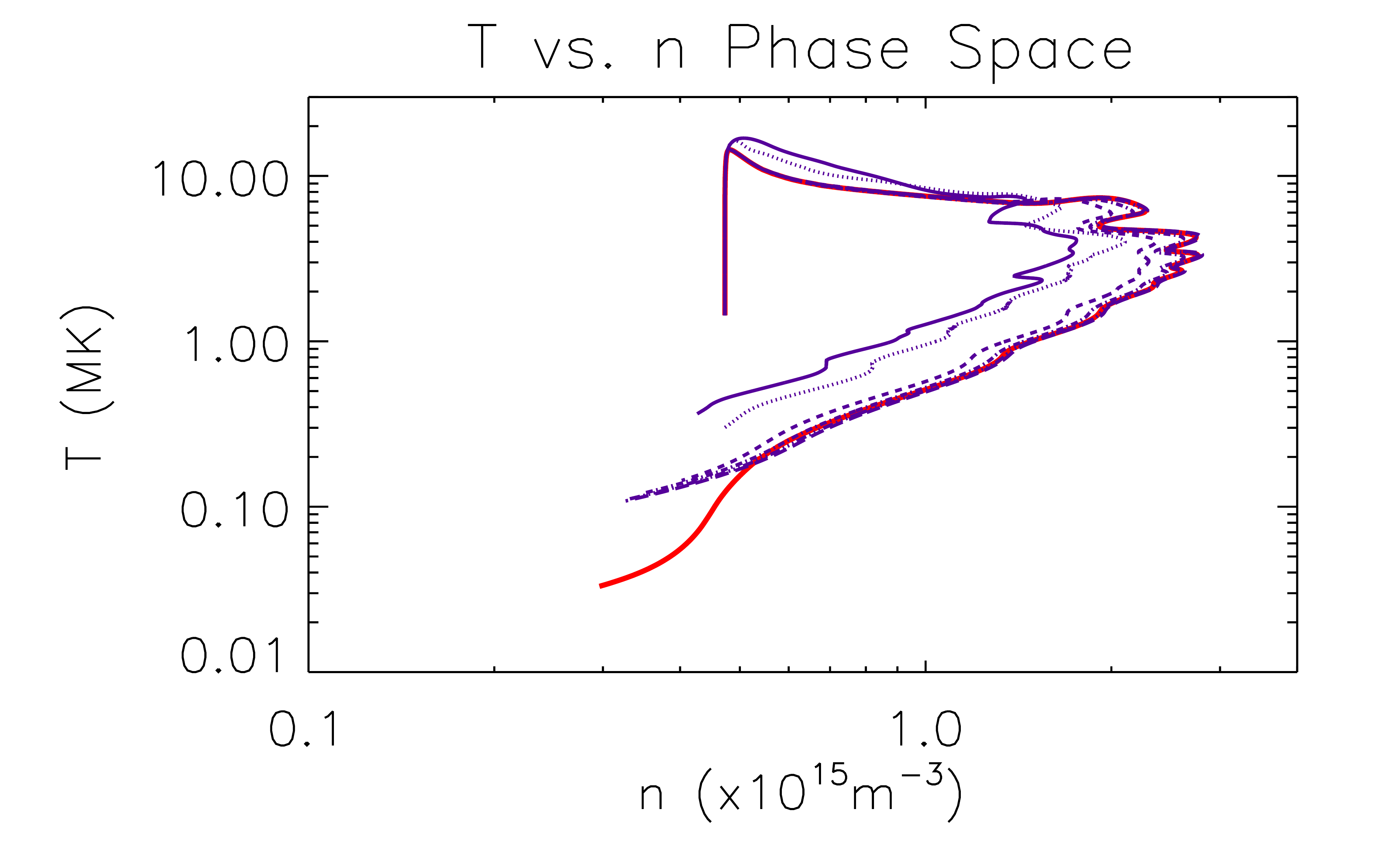}}
  \\
  \hspace*{-0.075\linewidth}
  \subfigure{\includegraphics[width=0.36\linewidth]
  {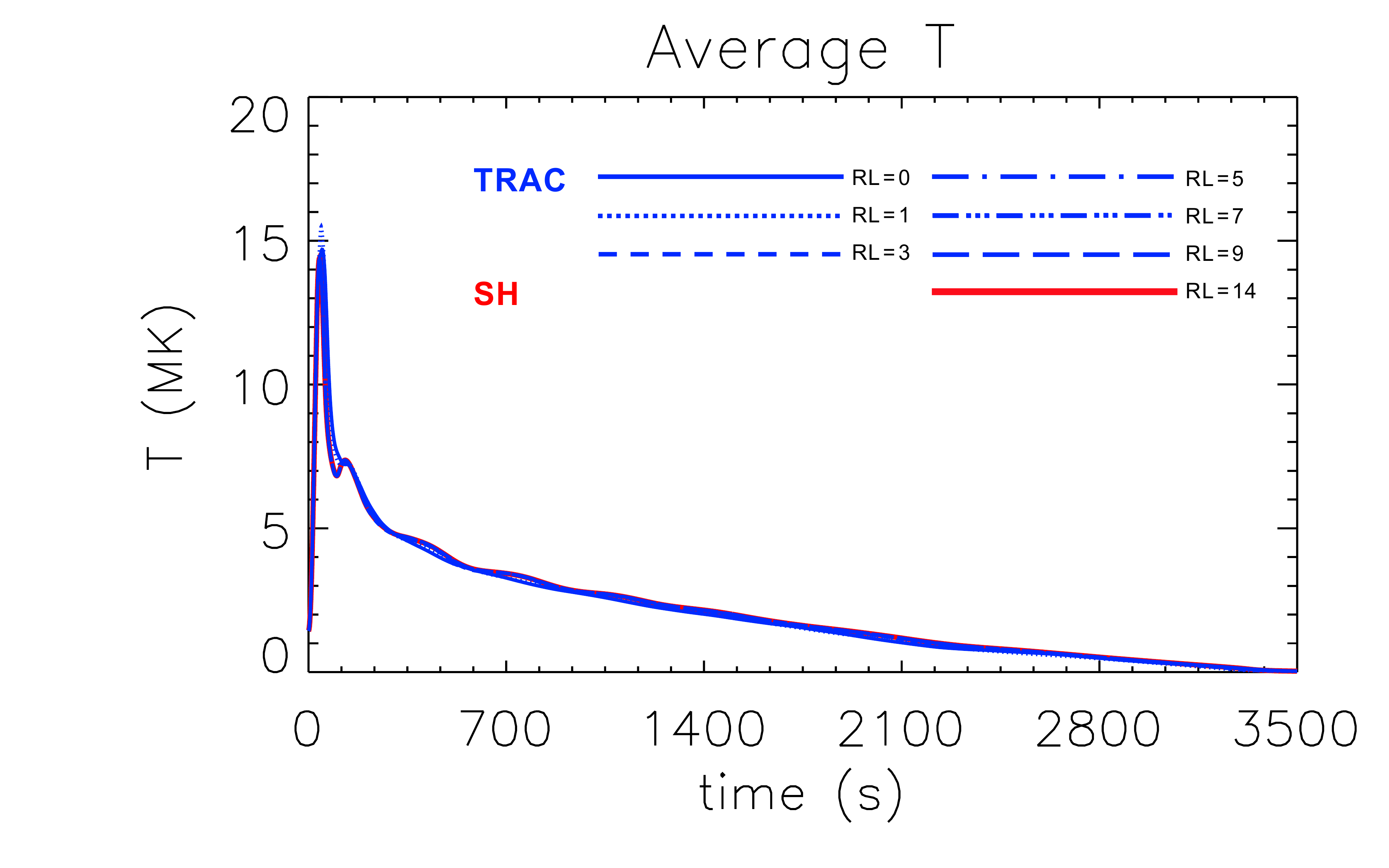}}
  \hspace*{-0.005\linewidth}
  \subfigure{\includegraphics[width=0.36\linewidth]
  {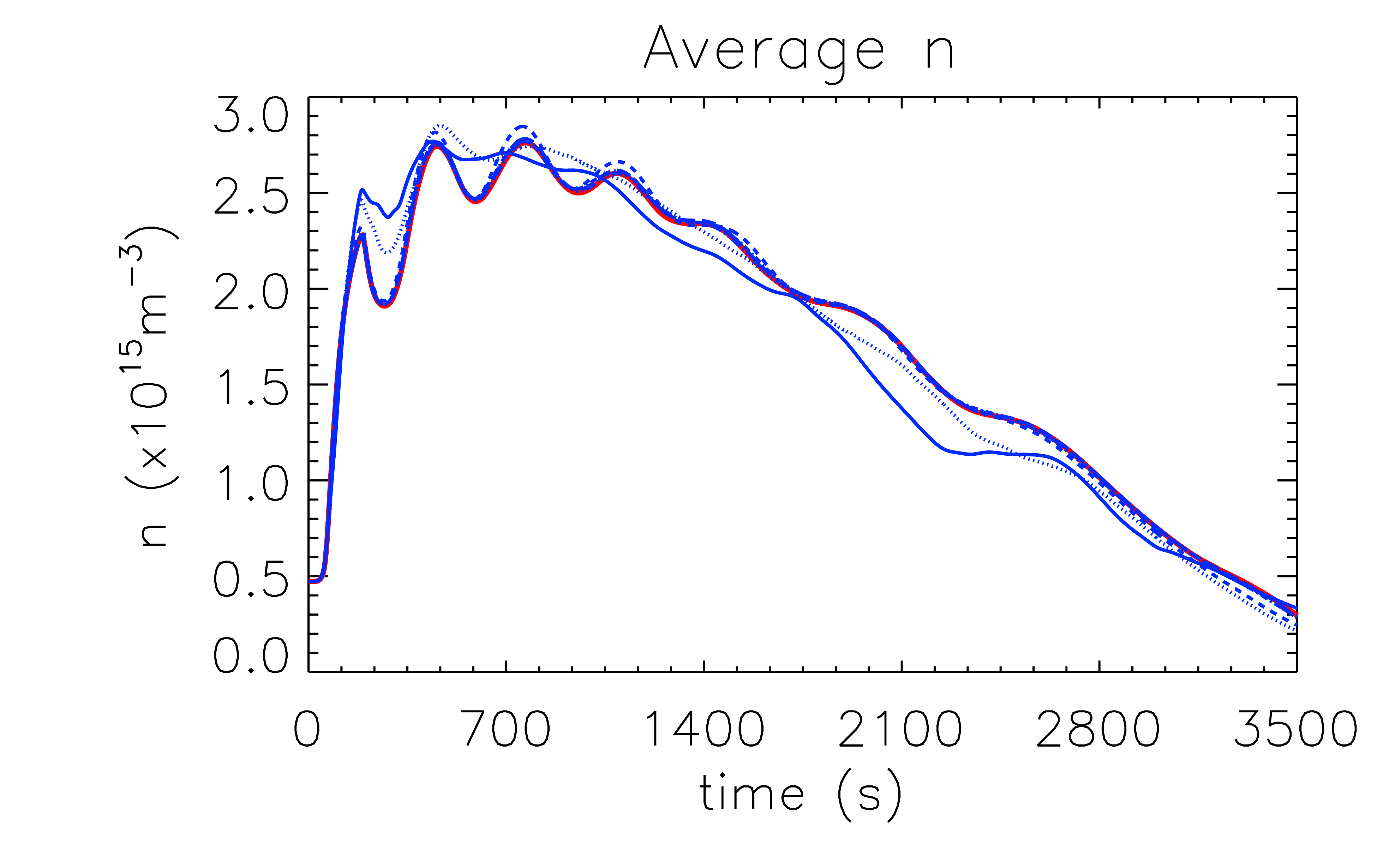}}
  \hspace*{-0.005\linewidth}
  \subfigure{\includegraphics[width=0.36\linewidth]
  {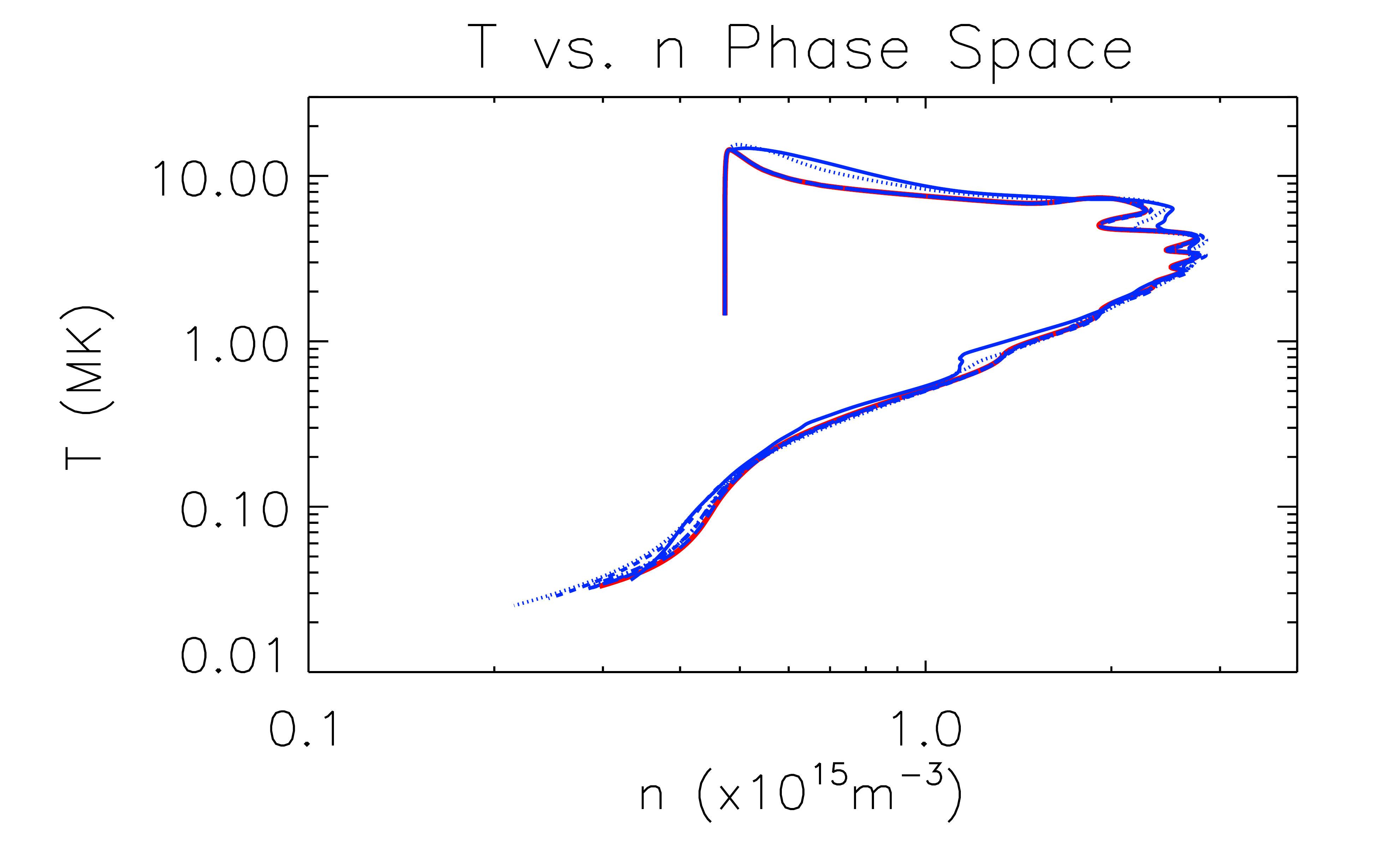}}
  \caption{Results for the 60 s heating pulse simulations. 
    The panels show the coronal
    averaged temperature (left-hand column) 
    and density (central column) as functions of time,
    and the temperature versus density phase space plot 
    (right-hand column). 
    The various curves represent different values of RL, which
    converge as RL increases (higher spatial resolution is 
    associated with larger RL).
    Rows 1--3 correspond to simulations run with the
    Spitzer-Harm (SH),
    \citet[][L09]{paper:Lionelloetal2009} and TRAC conduction 
    methods, respectively.
    The lines are colour-coded in a way that reflects the
    conduction method used. 
    Note that the properly resolved SH solution (RL=14) 
    is also
    shown in the L09 and TRAC panels.   
  \label{Fig:80Mm_60s_pulse_Tn_coronal_averages}
  }
\end{sidewaysfigure*}
  %
  %
  %
  %
%
%
  %
  %
\subsection{Validation Experiments
  \label{Sect:experiments}}
  \indent 
  The effectiveness of the TRAC method to obtain 
  the correct interaction between the corona and TR, 
  in response to rapid heating events, 
  is investigated by considering  
  two impulsive coronal heating events, 
  comprising short and long pulses that last
  for a total duration of 60 s and 600 s, 
  respectively.
  The temporal profile of the heating pulses is triangular
  with a peak value of 
  $Q_{\textrm{H}}=2\times 10^{-2}$ Jm$^{-3}$s$^{-1}$,
  while the spatial profile is uniform along the loop.
  We release the energy in a coronal loop of 
  total length 100 Mm, 
  which includes a 10 Mm chromosphere attached
  at the base of each TR. 
  Thus, the total energy injected into the coronal part
  of the loop is 
  $4.8 \times 10^7$ Jm$^{-2}$
  ($4.8 \times 10^8$ Jm$^{-2}$) 
  in the 60 s (600 s) heating pulse simulations.
  These heating conditions are representative of reasonably
  powerful flares
  and present a challenge to resolving the TR.
  \\
  \indent
  For each simulation, 
  the main assessment of the performance of
  TRAC is a comparison with the 
  results from two alternative
  methods that are commonly used to treat thermal conduction.
  Each method is
  applied 
  with spatial resolutions that cover several orders of
  magnitude, ranging from those  required for
  3D MHD codes to run in a realistic time to those 
  that fully resolve the TR but are achievable 
  only in field-aligned models.
  The first of these methods is the classical
  Spitzer-Harm heat flux formulation, while the second is 
  the approach developed by \cite{paper:Lionelloetal2009}, 
  which artificially broadens the TR below a 
  fixed specified 
  temperature. For brevity and clarity we define these two 
  conduction methods as 
  SH and L09, respectively.
  We note that the L09 method uses the broadening technique
  outlined in Section \ref{Sect:TRAC_broadening}
  but the modifications are
  applied below a fixed cutoff temperature 
  taken as $T_c=250,000$~K 
  (which is a typical value used by 
  \cite{paper:Lionelloetal2009} and
  \cite{paper:Mikicetal2013}).
  \\
  \indent
  Using the same set of parameter values but employing
  the three different conduction methods (SH, L09 and TRAC), 
  we repeated each simulation
  for RL = [0, 1, 3, 5, 7, 9, 11, 13, 14] refinement levels 
  to create a group of simulations 
  run for each conduction method. 
  \\
  \indent
  All of the simulations start from
  the same initial conditions. These are calculated using
  the SH parallel thermal conductivity and unmodified radiative
  loss rate.
  At $t=0$ (the initial conditions) a small 
  spatially uniform 
  background heating term ($Q_{\textrm{bg}}$) is
  present. This gives a starting temperature
  of order 1~MK. However, we note that  $Q_{\textrm{bg}}$
  is switched off thereafter and 
  the total energy in the initial conditions is 
  negligible compared with the energy released into the loop
  during the heating pulses.
  %
  %
  %
  %
\begin{sidewaysfigure*}
  \vspace{0.4\linewidth}
  \hspace*{-0.075\linewidth}
  \subfigure{\includegraphics[width=0.36\linewidth]
  {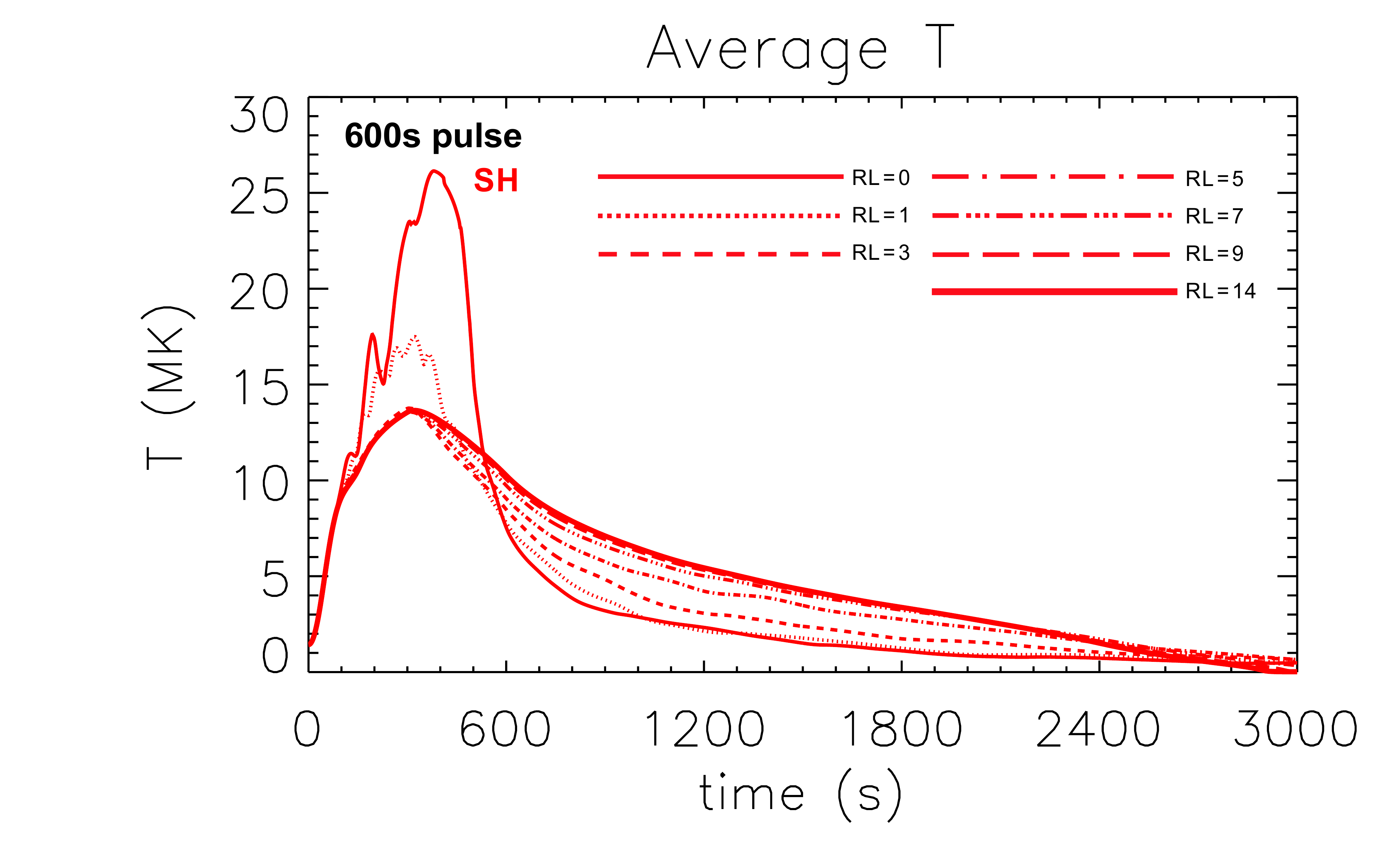}}
  \hspace*{-0.005\linewidth}
  \subfigure{\includegraphics[width=0.36\linewidth]
  {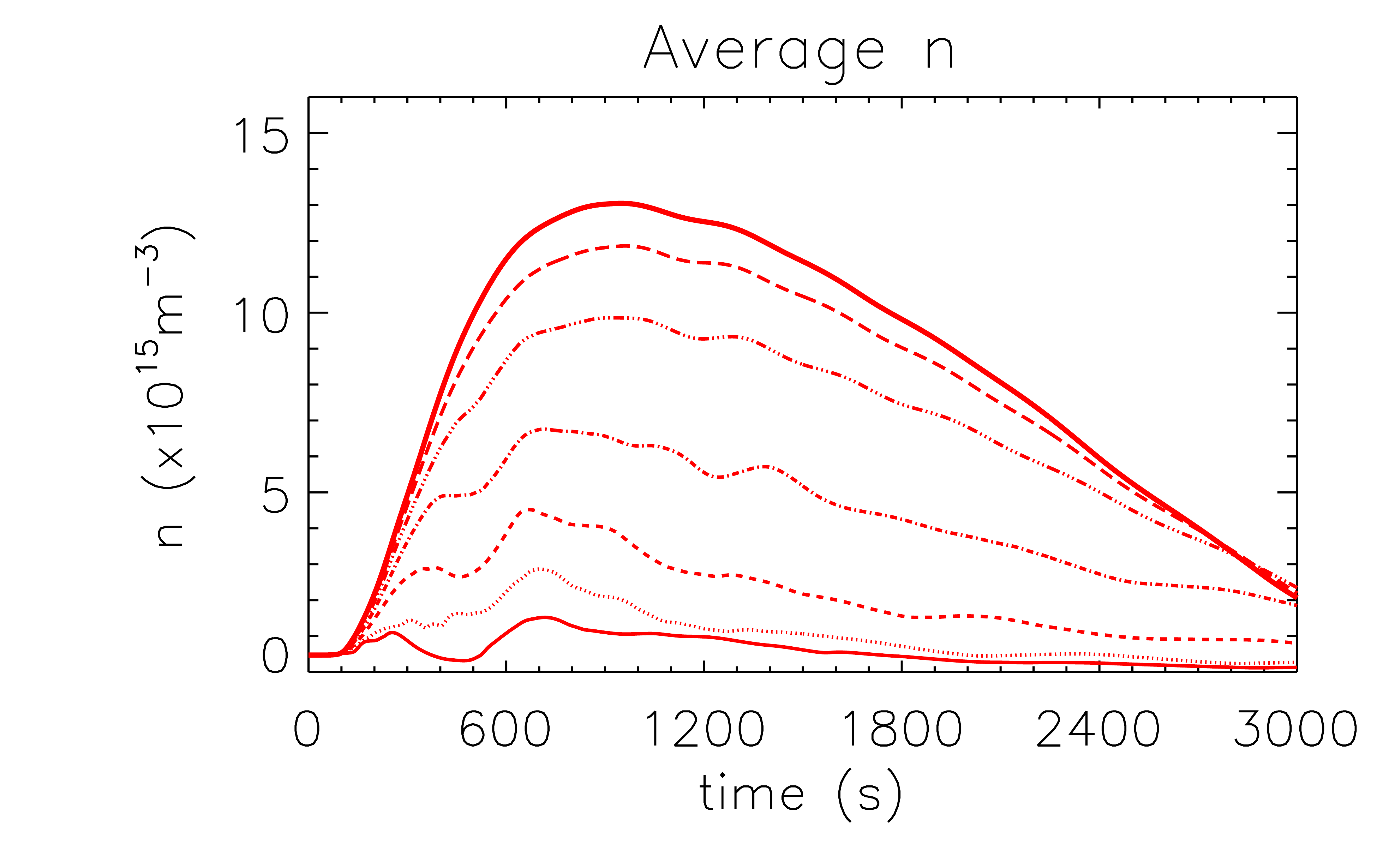}}
  \hspace*{-0.005\linewidth}
  \subfigure{\includegraphics[width=0.36\linewidth]
  {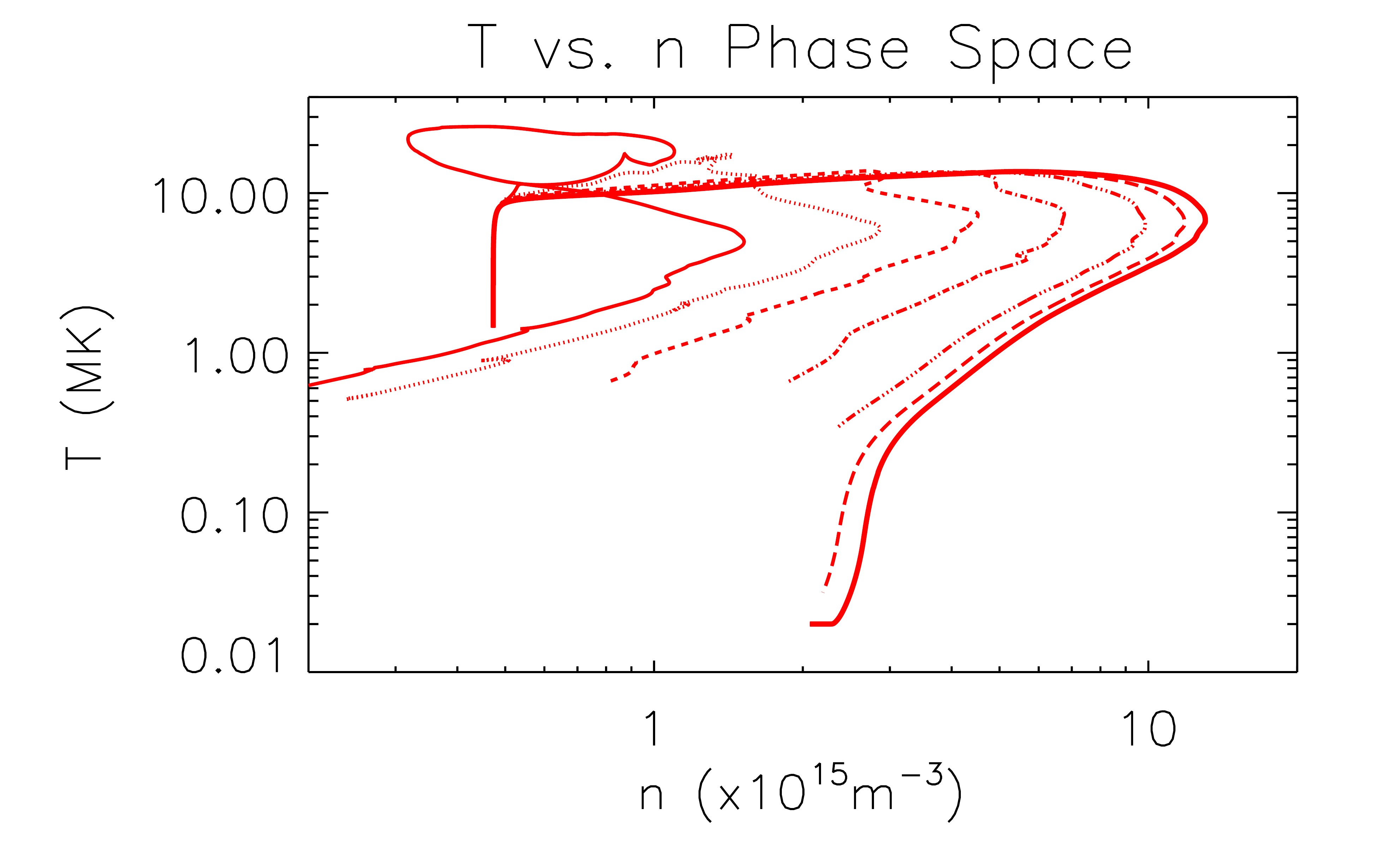}}
  \\
  \hspace*{-0.075\linewidth}
  \subfigure{\includegraphics[width=0.36\linewidth]
  {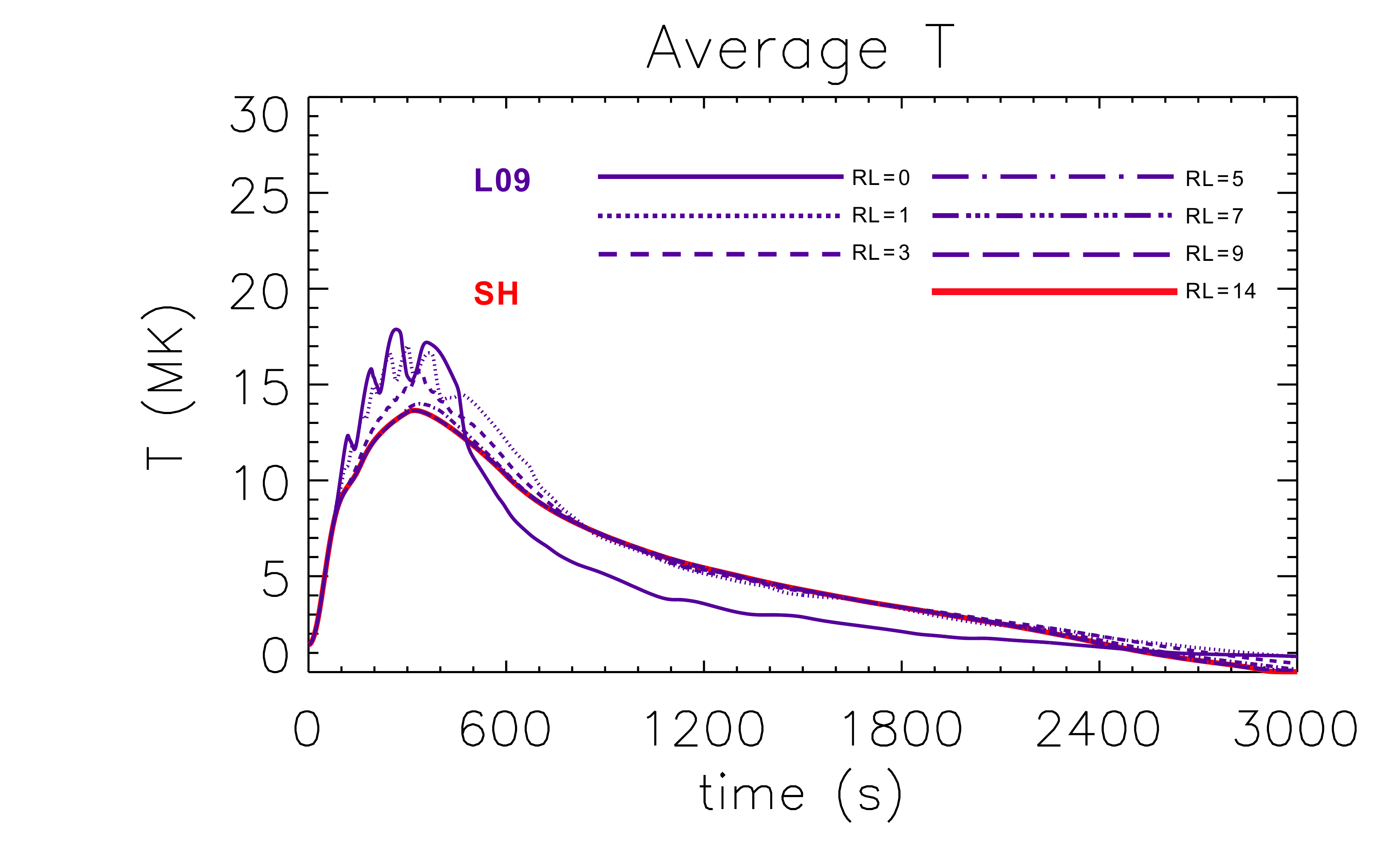}}
  \hspace*{-0.005\linewidth}
  \subfigure{\includegraphics[width=0.36\linewidth]
  {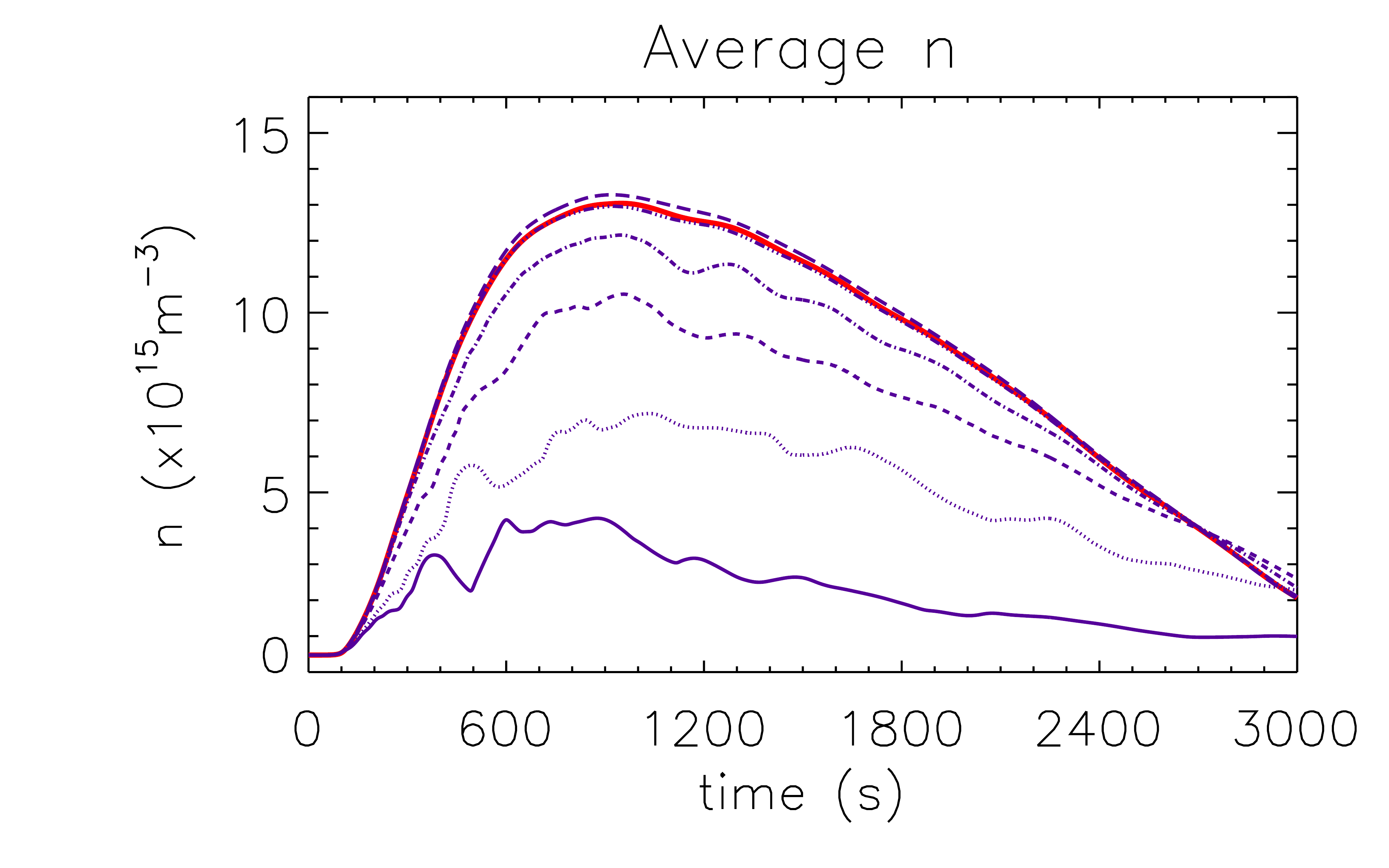}}
  \hspace*{-0.005\linewidth}
  \subfigure{\includegraphics[width=0.36\linewidth]
  {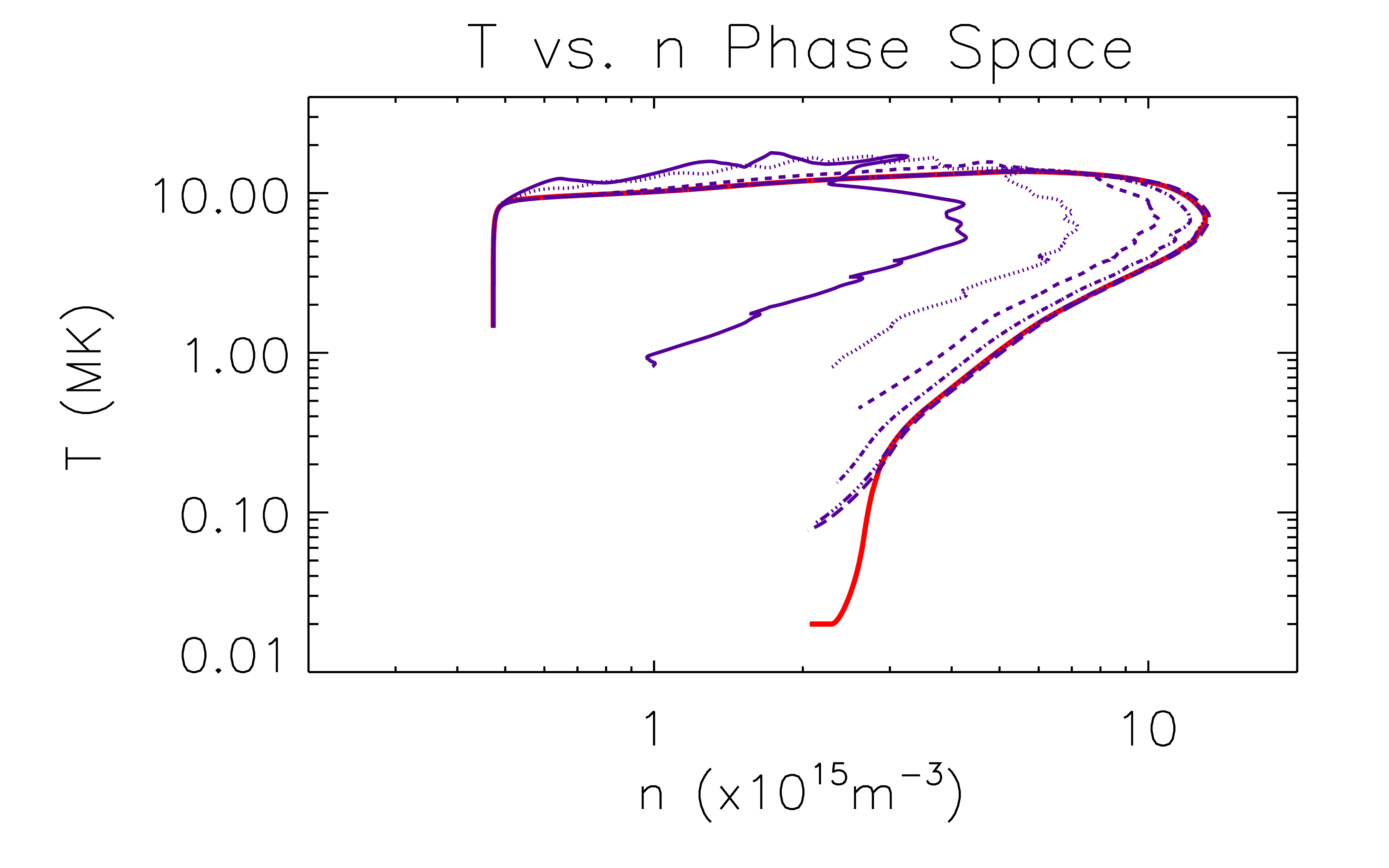}}
  \\
  \hspace*{-0.075\linewidth}
  \subfigure{\includegraphics[width=0.36\linewidth]
  {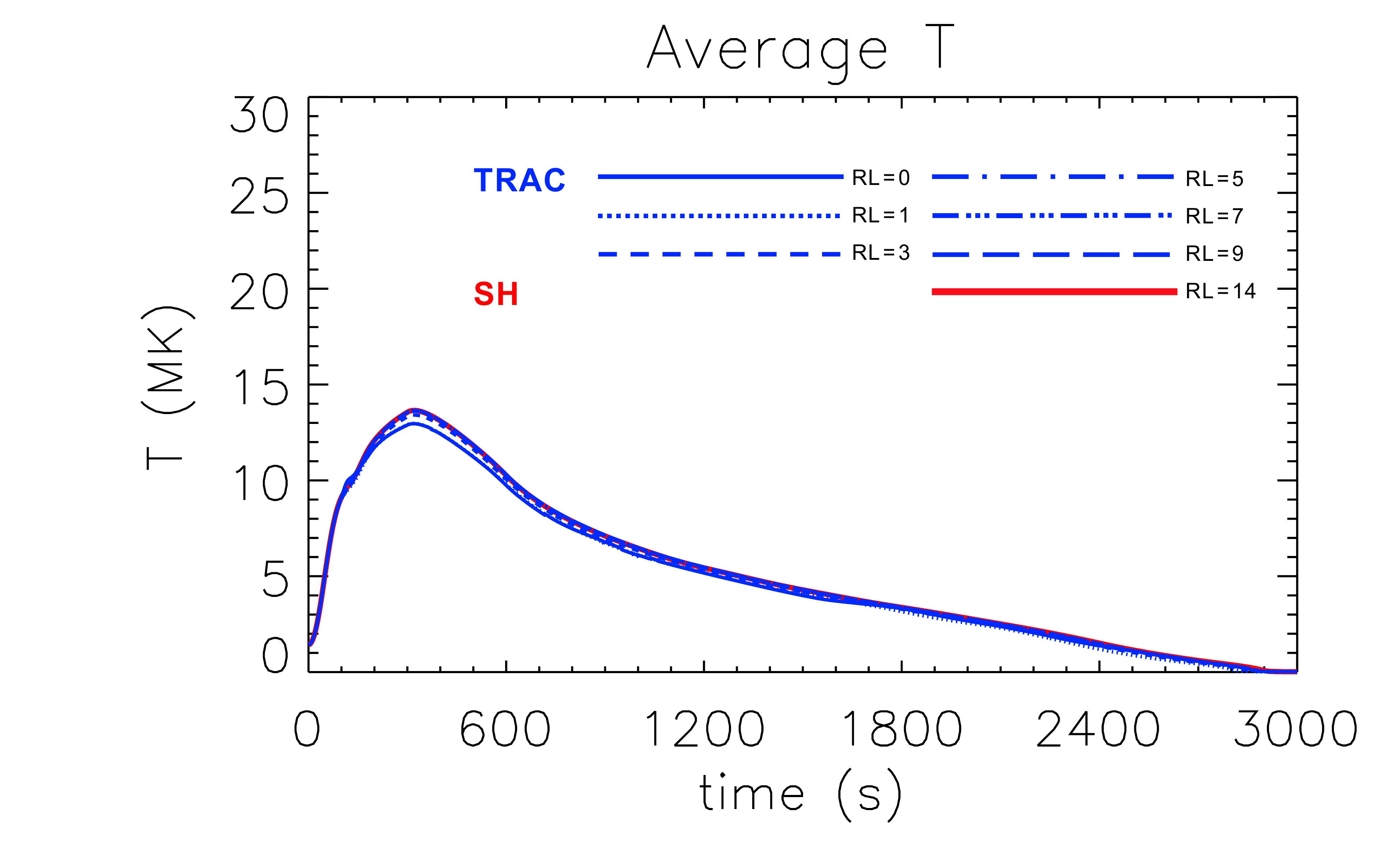}}
  \hspace*{-0.005\linewidth}
  \subfigure{\includegraphics[width=0.36\linewidth]
  {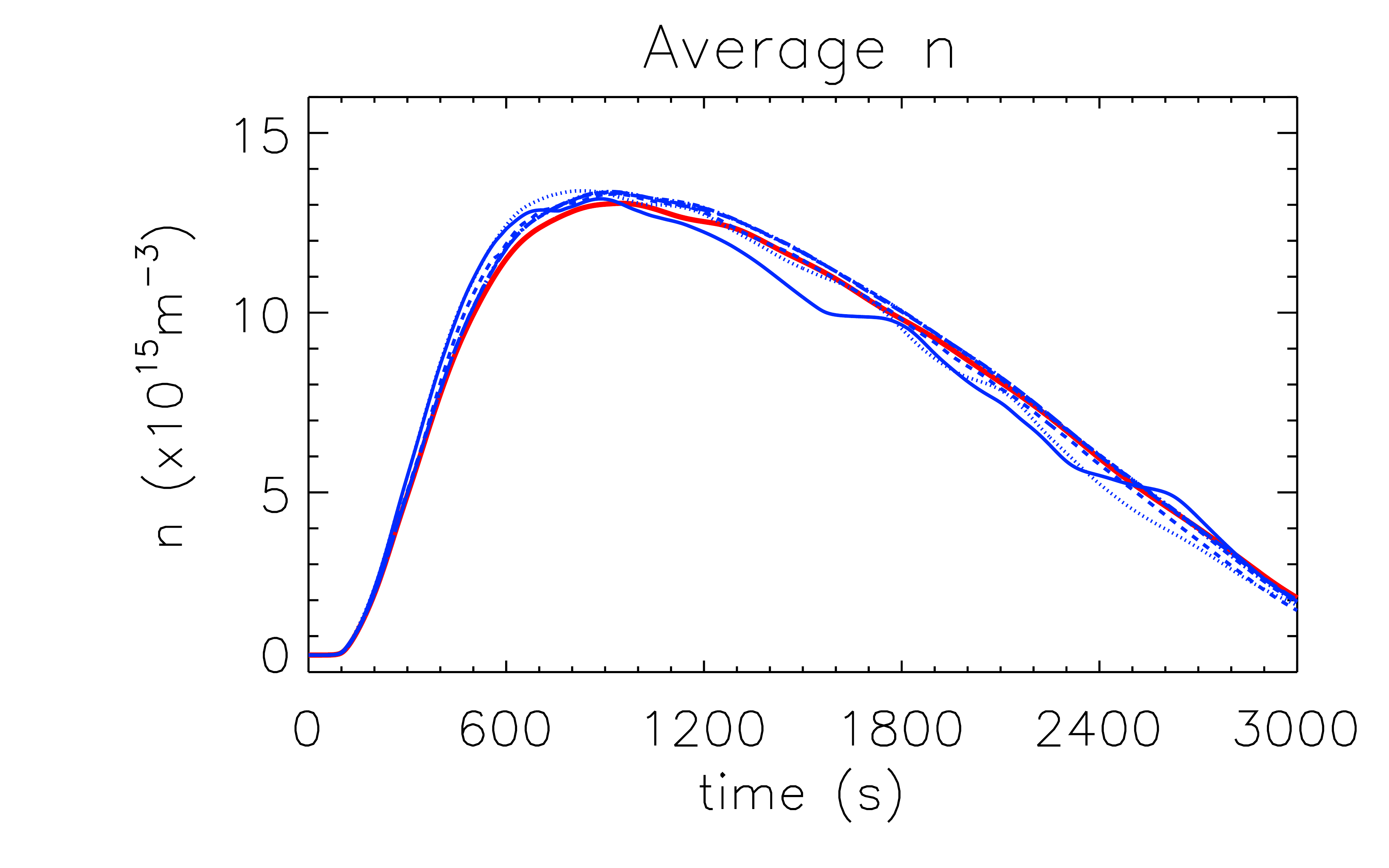}}
  \hspace*{-0.005\linewidth}
  \subfigure{\includegraphics[width=0.36\linewidth]
  {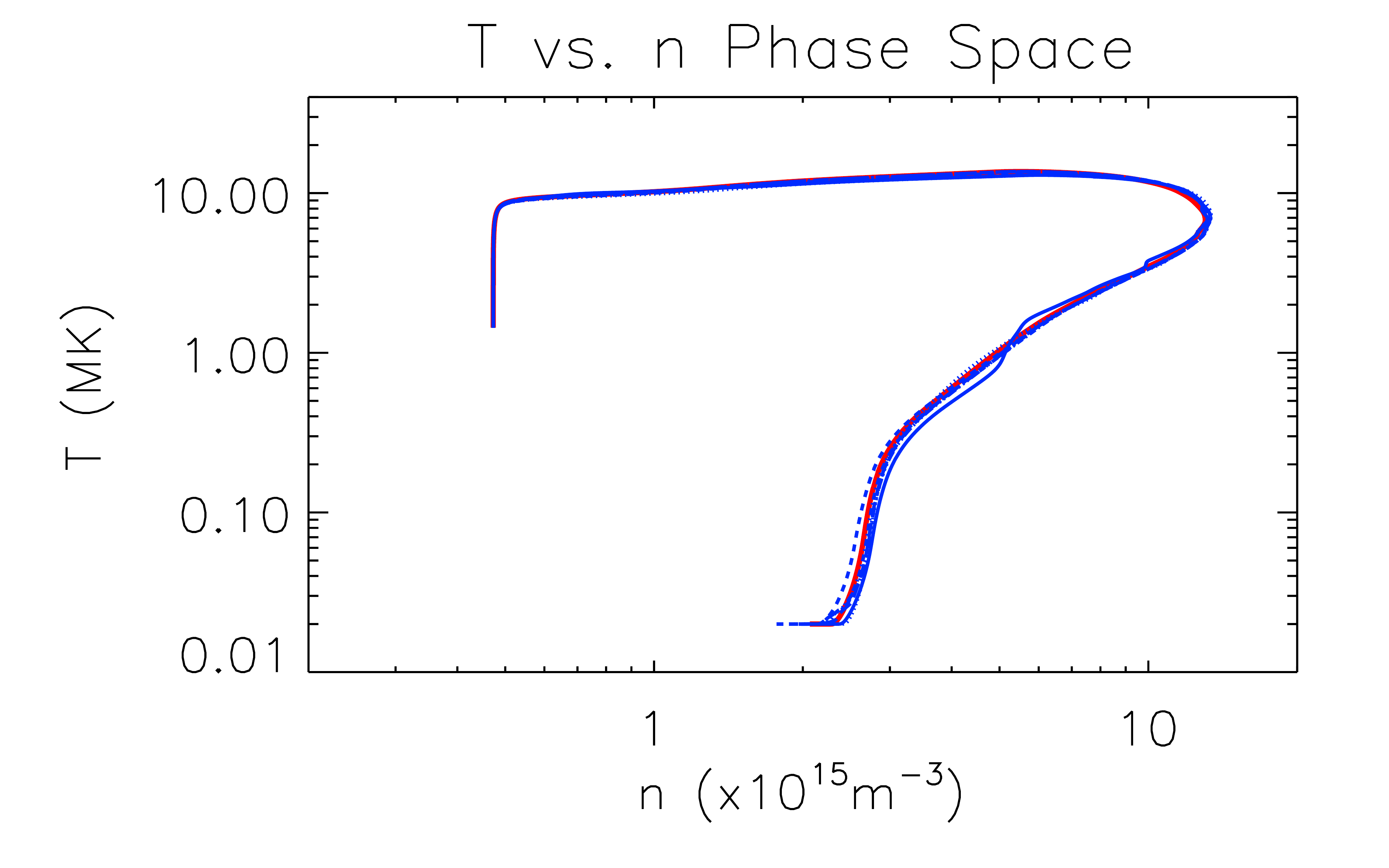}}
  \caption{Results for the 600 s heating pulse simulations.  
  Notation is the same as Figure
  \ref{Fig:80Mm_60s_pulse_Tn_coronal_averages}.    
  \label{Fig:80Mm_600s_pulse_Tn_coronal_averages}
  }
\end{sidewaysfigure*}
  %
  %
  %
  %
  %
  %
\begin{figure*}
  \hspace*{-0.02\linewidth}
  \subfigure{\includegraphics[width=0.5\linewidth]
  {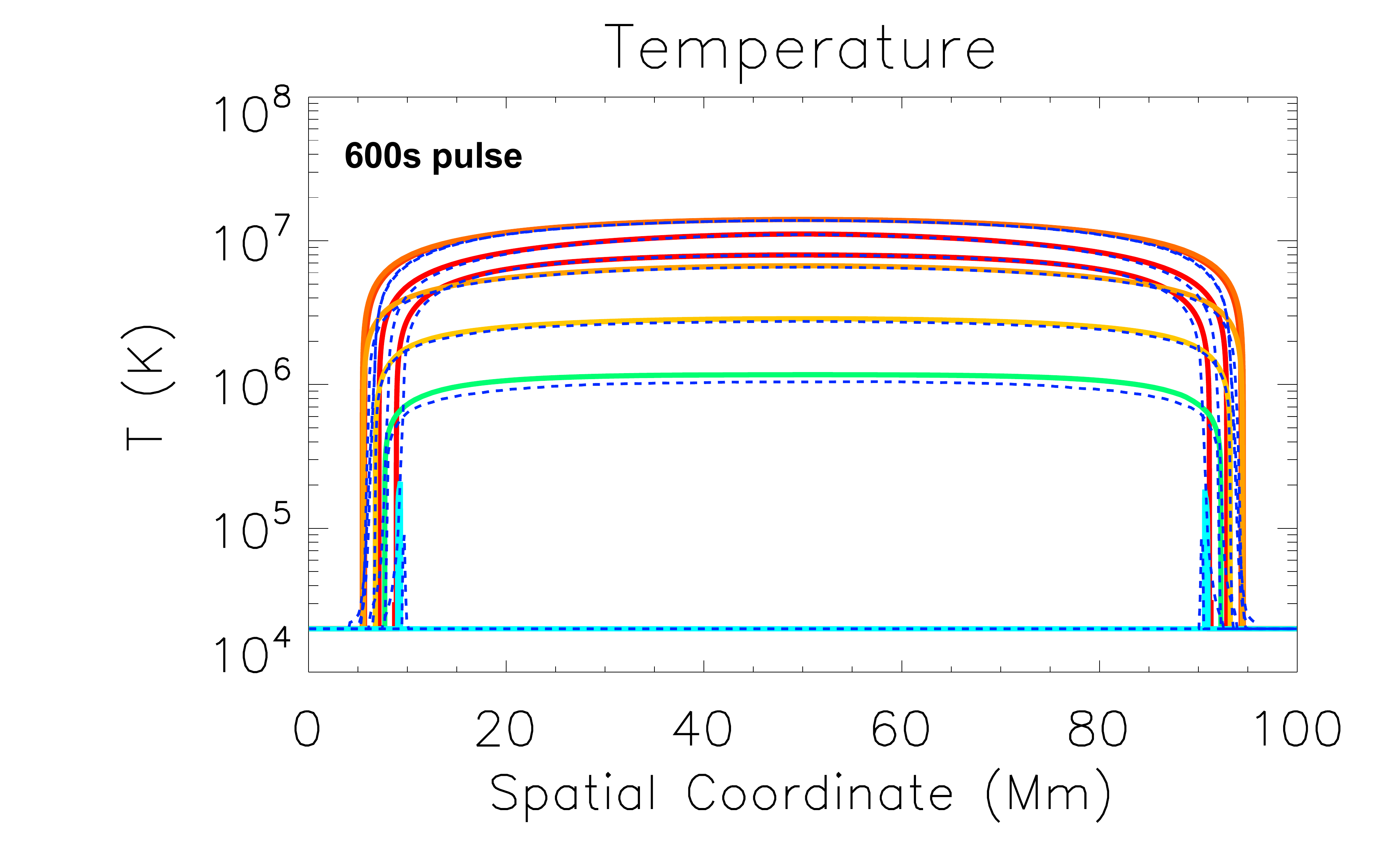}}
  \subfigure{\includegraphics[width=0.5\linewidth]
  {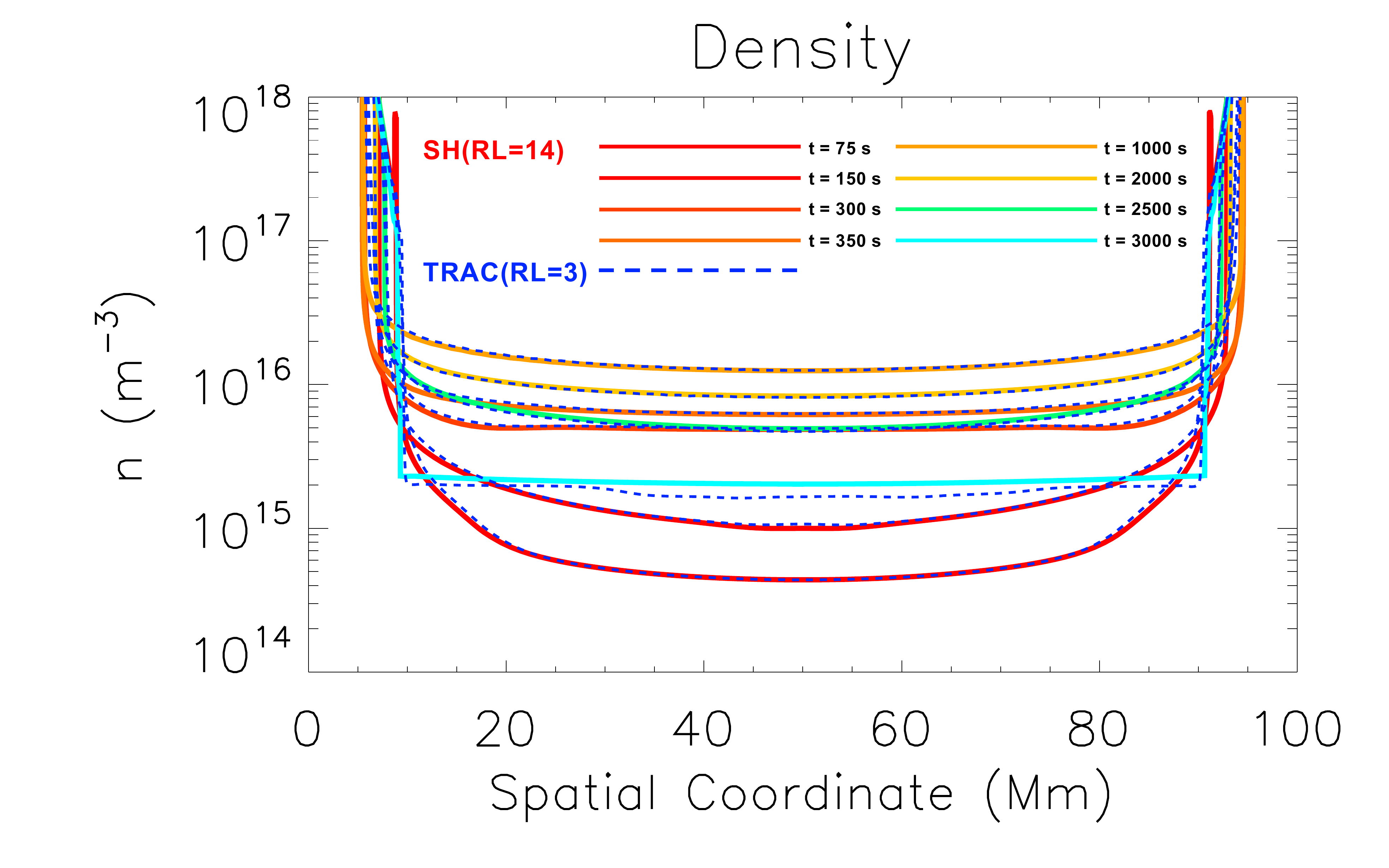}}
  \\[-5mm]
  \hspace*{-0.02\linewidth}
  \subfigure{\includegraphics[width=0.5\linewidth]
  {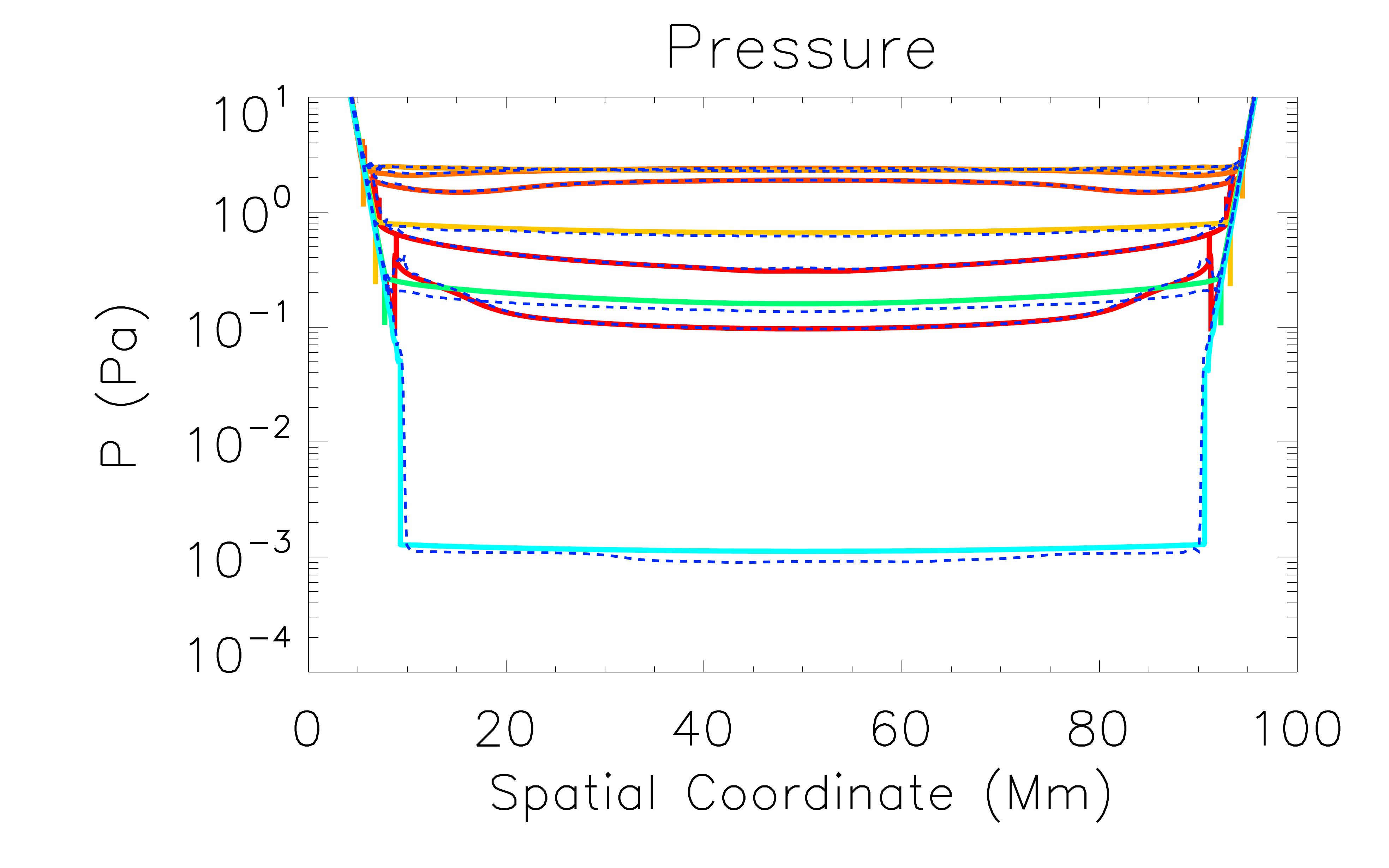}}
  \subfigure{\includegraphics[width=0.5\linewidth]
  {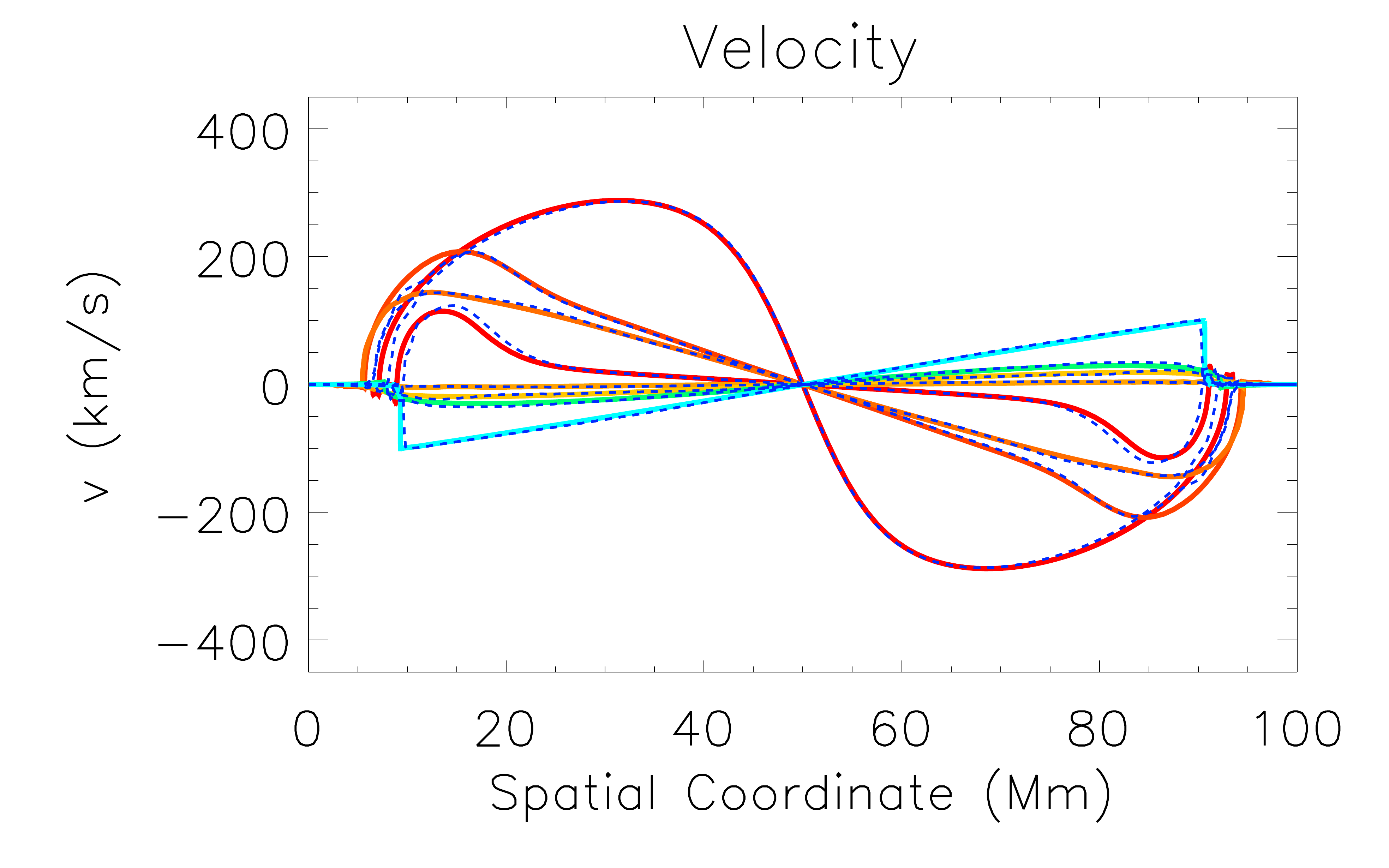}}
  \\[-2.5mm]
  \hspace*{-0.02\linewidth}
  \subfigure{\includegraphics[width=0.5\linewidth]
  {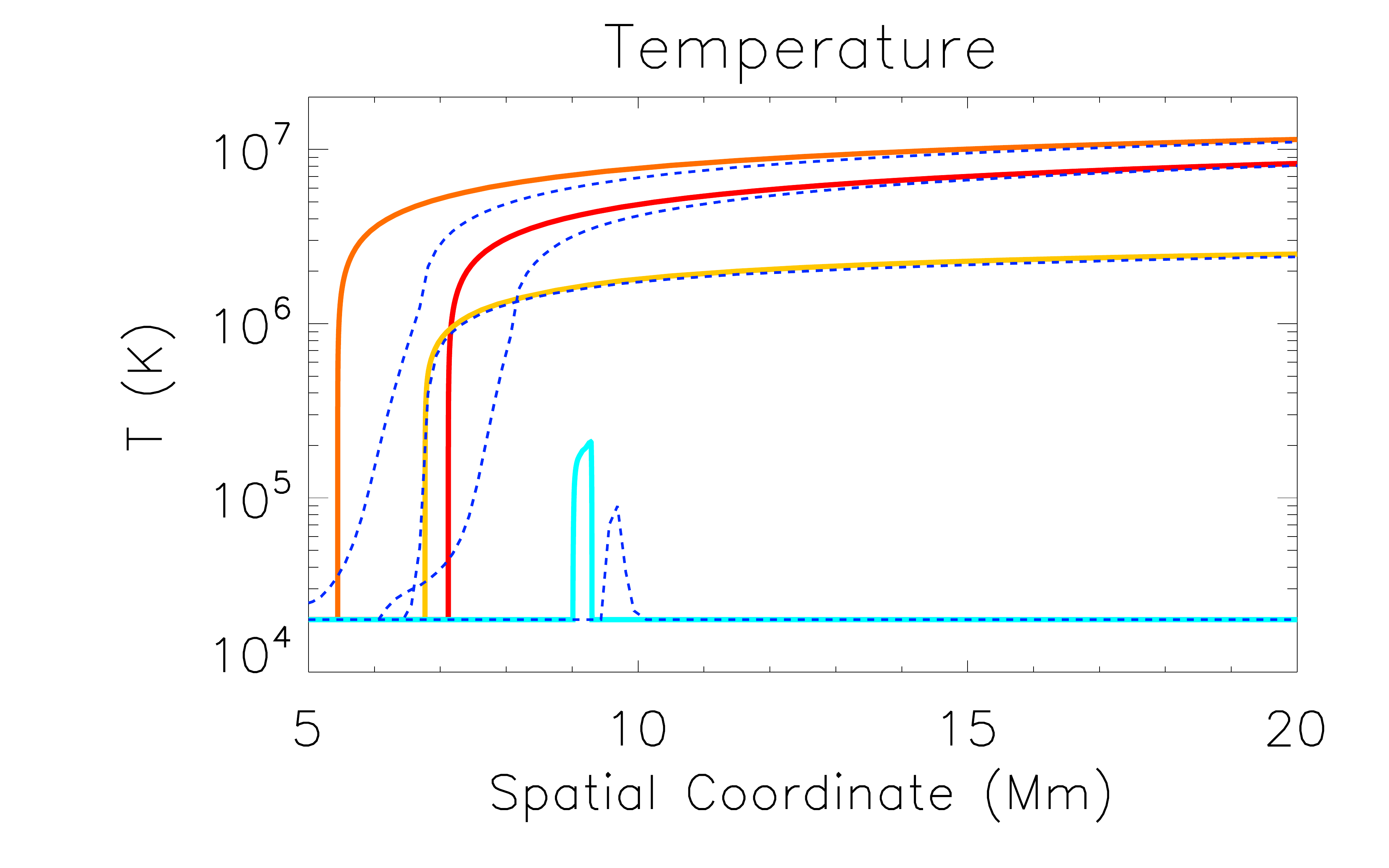}}
  \subfigure{\includegraphics[width=0.5\linewidth]
  {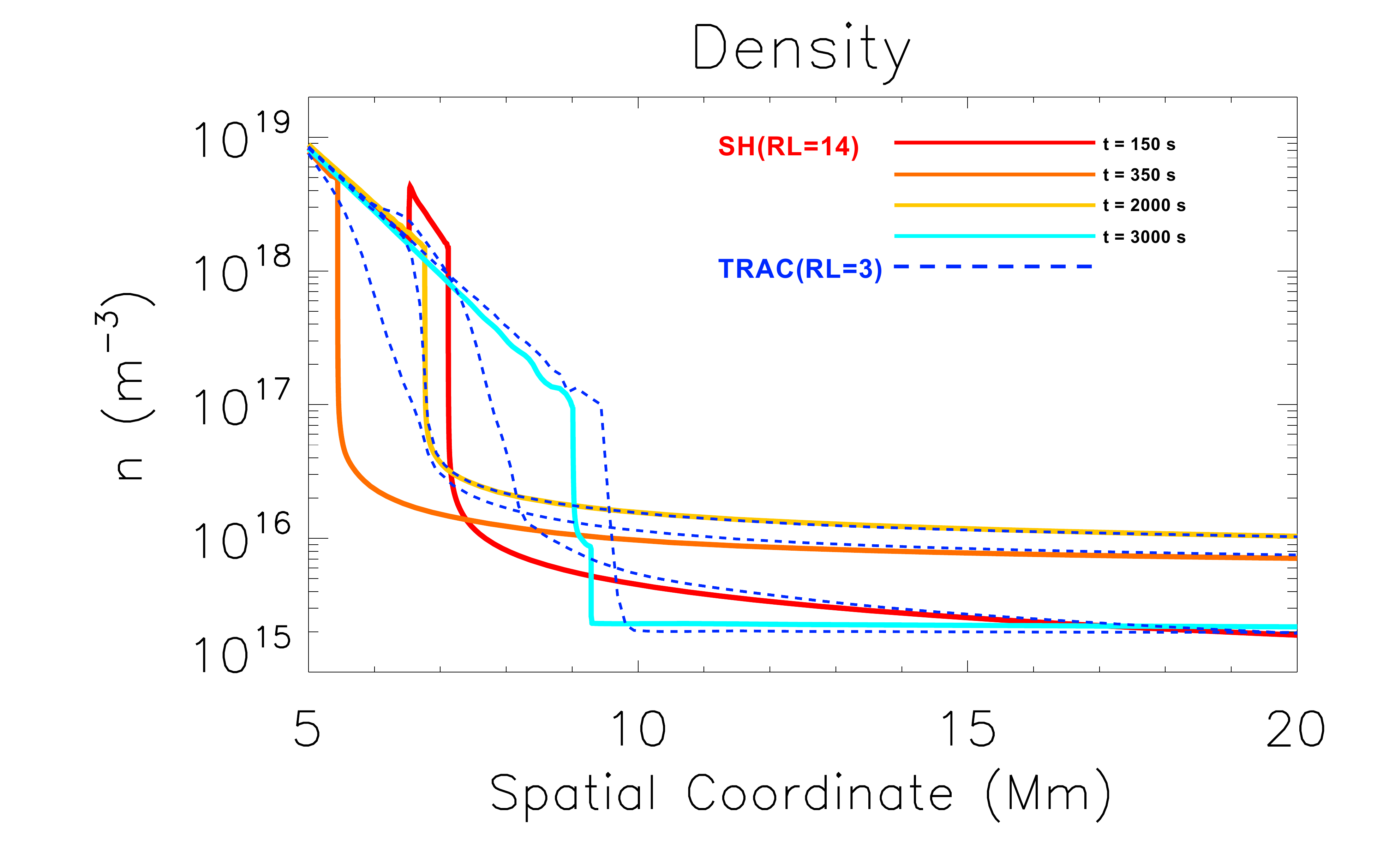}}
  \\[-5mm]
  \hspace*{-0.02\linewidth}
  \subfigure{\includegraphics[width=0.5\linewidth]
  {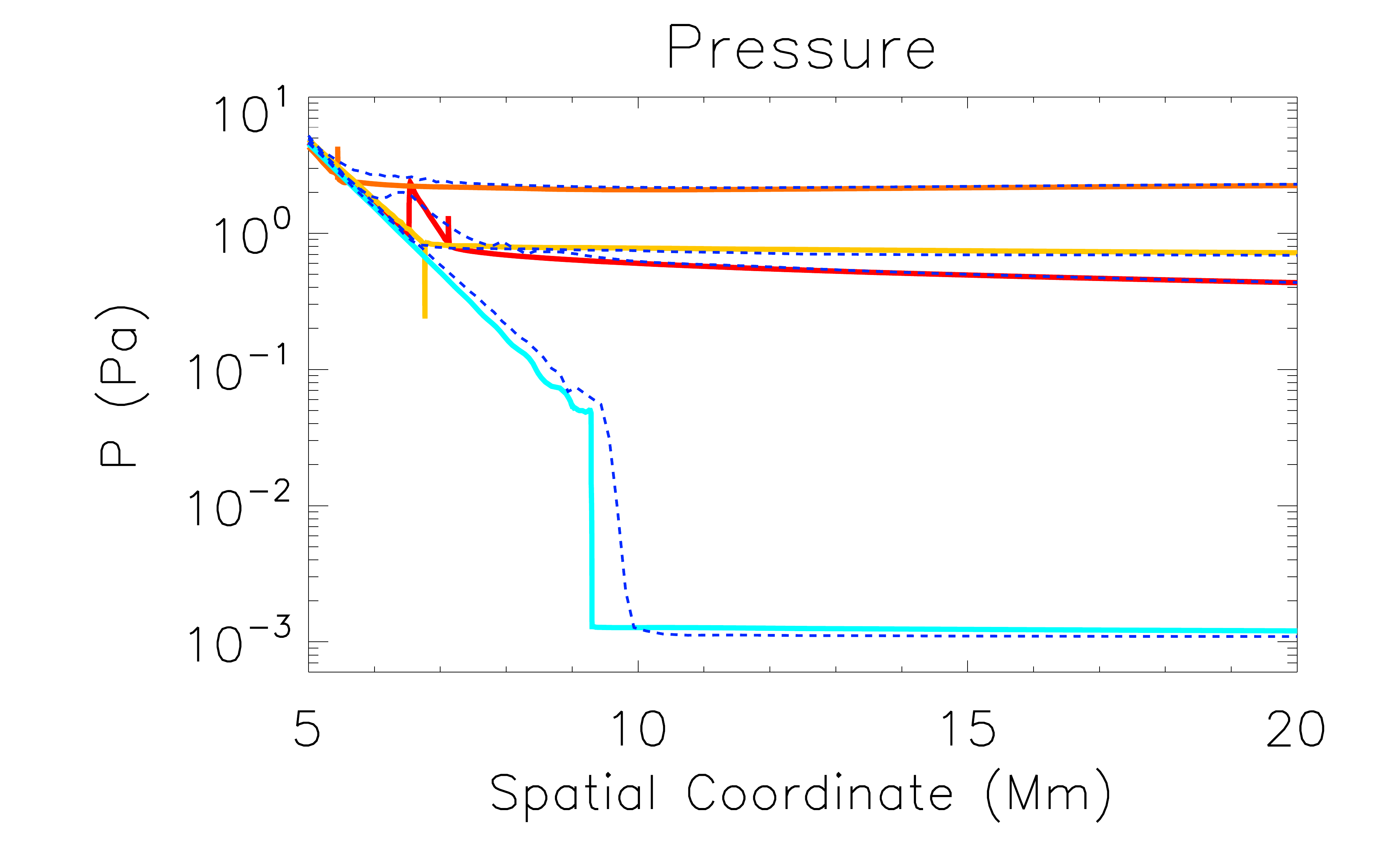}}
  \subfigure{\includegraphics[width=0.5\linewidth]
  {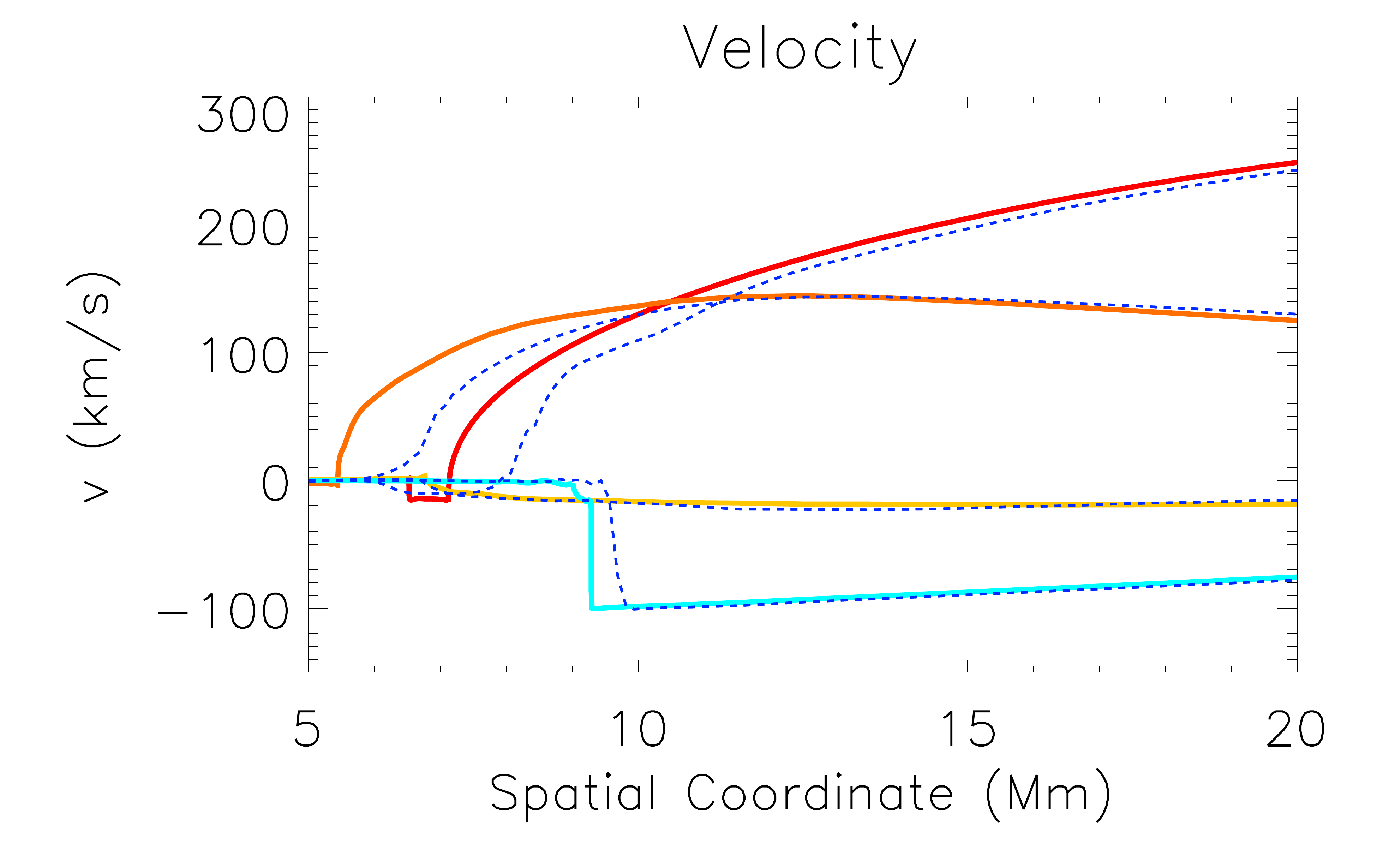}}
  \caption{Results for the 600 s heating pulse simulations. 
    Starting from the top left, the upper four 
    panels show time ordered
    snapshots of the temperature, density, pressure and 
    velocity as functions of position along the loop 
    for times during the ablation and decay phases.  
    The lower four panels 
    show an enlargement about the 
    transition region of a subset of these snapshots.
    The various solid curves represent 
    the properly resolved SH solution (RL=14) at 
    different times (with
    the lines colour-coded in a way that reflects the
    temporal evolution) and the dashed blue curves correspond 
    to the TRAC solution that is computed with RL=3 at these 
    times. 
    Animations of the 60~s and 600~s heating pulse simulations
    can be viewed in the online version of 
    this article.
    These compare the full time evolution of the 
    $T, n, p$ and $v$ 
    profiles of the SH simulations run with 
    RL=14 and the TRAC simulations run with RL=3.
  \label{Fig:80Mm_600s_pulse_global_Tnpv}
  }
\end{figure*}
  %
  %
  %
  %
%
%
  %
  %
\section{Results
  \label{Sect:res}}
   %
   %
%
%
  %
  %
\subsection{Coronal Response to Heating
  \label{Sect:coronal_response}}
  \indent
  Figure \ref{Fig:80Mm_60s_pulse_Tn_coronal_averages}
  shows the temporal evolution of the coronal
  averaged temperature ($T$) and density ($n$)
  and the corresponding $T$ versus $n$
  phase space plot (shown on a log-log scale) 
  for the 60 s heating pulse simulation.
  The coronal averages are calculated by spatially
  averaging over the uppermost 50\% of the loop.
  Each conduction method is identified with a specific colour
  and their results are shown in separate rows. 
  The red curves correspond to the SH method (first row), 
  the purple curves the L09 method (second row) 
  and the blue curves represent
  the TRAC method (third row). 
  \\
  \indent
  In the panels of each method, each curve corresponds to
  a simulation run with a 
  different value of maximum refinement level. 
  Taking the SH simulations as an example,
  the line styles associated with the different
  refinement levels are shown in the figure legend
  on the temperature plot, and indicate 
  that RL increases as one moves
  upwards from the lowest curve (RL=0) 
  to the highest curve (RL=14)
  in the density 
  plot.
  These simulations are identical in all respects
  except for the value of RL. 
  Note that the results
  for RL=11 and 13 are not shown in the SH panels.
  In addition, 
  RL=11, 13 and 14 are not shown in the L09 and TRAC
  panels, where instead  
  the SH simulation run with RL=14 (solid red curves) 
  is shown.
  This SH simulation (RL=14) is
  properly resolved and used as a benchmark solution.
  \\
  \indent
  The coronal response of the benchmark simulation
  (SH with RL=14), to the 60 s heating pulse, 
  follows the
  standard evolution of an impulsively heated loop
  as described in the literature
  \citep{paper:Bradshaw&Cargill2006,
  paper:Cargilletal2012a,paper:Cargilletal2012b,
  paper:Cargilletal2015,
  paper:Bradshaw&Klimchuk2015,
  paper:Reale2016}. 
  The rise in temperature is
  followed by the density increasing due to the
  ablation process
  \citep{paper:Antiochos&Sturrock1978,
  paper:Klimchuketal2008},
  then, after the time of the density peak, 
  the loop cools by radiation and drains by a downward 
  enthalpy flux to the TR
  \citep{paper:Bradshaw&Cargill2010a,
  paper:Bradshaw&Cargill2010b}.
  The temperature cools below the 
  starting value of 1~MK at around 2100s, and then
  continues to decrease towards the chromospheric value
  of $2\times 10^4$~K 
  while the density is evacuated from the loop.
  \\
  \indent
  Consistent with BC13, the coronal density evolution 
  in the group of 
  SH simulations is strongly dependent on the
  spatial resolution, 
  requiring grid cell widths of at least 488 m 
  (RL$\geq11$)  for convergence.
  These resolution requirements are 
  slightly weaker in the L09 simulations, 
  with convergence seen for minimum
  grid cell widths of 7.81 km or 
  finer (RL$\geq7$).  
  However, the L09 method
  achieves this 
  relatively modest relaxation of the resolution
  dependence
  at the expense of accurately 
  modelling 
  the loop evolution below the 
  specified
  cutoff temperature ($T_c=250,000$~K).
  \\
  \indent
  It is therefore striking that 
  the TRAC solutions are only weakly
  dependent on the spatial resolution, while maintaining 
  high levels of accuracy throughout the full coronal
  response range.
  Grid cell widths of 125 km (RL=3, medium dashed curves)  
  are sufficient to observe
  convergence
  to the properly resolved SH solution which employed
  TR grid cell widths of 61 m.
  The TRAC solution computed with RL=3
  correctly captures the interaction between the corona
  and chromosphere during the ablation phase, up to
  the time of peak density
  where the error is less than 3\% and the subsequent decay 
  phase.
  Even the density oscillations 
  (that are characteristic of the short heating pulse imposed 
  e.g \cite{paper:Reale2016})
  and the global cooling of the loop 
  down to $2\times 10^4$~K,
  are both correctly captured.
  This demonstrates that adaptively broadening unresolved parts
  of the TR
  removes the influence of numerical resolution on the 
  coronal density response to impulsive heating.
  \\
  \indent 
  Figure \ref{Fig:80Mm_600s_pulse_Tn_coronal_averages}
  shows the results for the 600 s heating pulse simulations.
  Releasing the energy for an extended period of time but
  with the same peak heating rate
  has two main consequences. 
  Firstly, 
  the density oscillations 
  associated with the sloshing of the
  coronal plasma disappear.
  Secondly, 
  the differences between the density curves
  that represent the various SH
  and L09 simulations become more severe.
  \\
  \indent
  Considering 
  the refinement level of RL=3 as an example,
  these conduction methods have peak density 
  errors of 70\% (SH) and 20\% (L09)
  when benchmarked against 
  the properly resolved simulation 
  (SH with RL=14),
  increased from
  discrepancies of 55\% (SH) and 15\% (L09) 
  for the shorter heating pulse.
  Consequently, the convergence 
  requirement on the spatial
  resolution
  becomes stricter
  for the L09 method and remains as severe for the
  SH method.
  Grid cell widths of 488 m (RL$\geq 11$)
  and 1.95 km (RL$\geq 9$) were necessary for the 
  solutions to converge in the SH and L09 
  simulations, respectively.
  \\
  \indent
  On the other hand, the TRAC solutions once again converge
  at RL=3 (125 km grid cells) with a
  peak density error of less than 2\%,  
  indicating that this treatment of the TR 
  gives rise to a coronal response that
  can accurately follow the complete coronal heating and 
  cooling cycle with coarse spatial resolutions, irrespective
  of the heating duration and resulting density regime.
  \\
  \indent
  In particular, TRAC eliminates the need for the
  very short time steps that are imposed
  by a highly resolved numerical grid.
  This enables the method to accurately capture the corona/TR
  enthalpy exchange significantly faster than fully resolved 
  models.
  A typical speed-up is between two and three orders of 
  magnitude.
  For example, the computation time of the properly resolved SH
  simulation (RL=14) is 3.5 days for the 600~s heating pulse,  
  while the run time for the TRAC simulation computed with RL=3
  is only 6 minutes.
  %
  %
%
%
  %
  %
\subsection{Global Evolution
  \label{Sect:global_evolution}}
  \indent
  The main question that needs to be addressed in understanding
  these results is why are the TRAC simulations 
  so successful in describing the 
  coronal response to heating
  with such large grid cell widths?
  A comprehensive answer requires a detailed assessment
  of how the TRAC modifications to the 
  parallel thermal conductivity and radiative loss rate
  affect the local energy balance and
  subsequent dynamics in the TR, coupled 
  together with the resulting global
  evolution of the loop.
  Here we concentrate just on the latter. 
  The former
  will be presented in future work.
  \\
  \indent
  Figure
  \ref{Fig:80Mm_600s_pulse_global_Tnpv}
  shows the time evolution of the
  global temperature, 
  density, pressure, and velocity profiles 
  for the 600 s
  heating pulse simulations.
  Each solid curve represents a
  different snapshot from
  the properly resolved SH simulation (RL=14)
  and the dashed blue curves imposed on top
  are the corresponding
  snapshots from the TRAC simulation computed with
  125 km grid cells (RL=3).
  (The comparison with an under-resolved SH solution
  (e.g.~SH run with RL=3)
  has been previously discussed in BC13 and 
  \cite{paper:Johnstonetal2017a} and will not be 
  considered further.)
  \\
  \indent
  First we focus on the evolution of the pressure.
  Of particular importance is the formation of the
  pressure gradients in the TR.
  Adaptively broadening the temperature and density profiles
  in the unresolved parts of the TR
  prevents the downward 
  heat flux jumping across the unresolved region.
  This ensures that the incoming energy
  goes into increasing the gas pressure locally, 
  rather than being lost due to artificially high 
  radiative losses
  \citep[e.g.~BC13;][]
  {paper:Johnstonetal2017a,paper:Johnstonetal2017b}.
  The scale of the broadening is relatively small 
  but Figure \ref{Fig:80Mm_600s_pulse_global_Tnpv} 
  shows that the resulting
  TRAC approximations of the SH pressure gradients
  are remarkably good. 
  These pressure gradients 
  are then  responsible for driving the flows.
  \\
  \indent
  To that end
  it is clear the TRAC solution
  correctly captures the global velocity evolution of 
  the properly resolved SH simulation,
  during both the
  ablation and decay phases in response to the 600~s heating 
  pulses. 
  (The 60~s heating pulse simulations show the same
  fundamental properties).
  This ensures that TRAC captures the mass 
  and energy exchange 
  that takes place
  between the chromosphere, TR and corona correctly,
  enabling the method to maintain high levels of 
  accuracy in the coronal temperature and density
  evolution
  throughout the full heating and upflow followed by
  cooling and downflow cycle.
   %
   %
%
%
  %
  %
  \section{Discussion}
  \label{Sect:dis_concl}
  \indent
  The difficulty of obtaining adequate spatial resolution
  in numerical models of the outer solar atmosphere has been
  a long-standing problem.
  BC13 demonstrated that lack of adequate spatial
  resolution during impulsive heating events
  led to coronal densities that are erroneously small. 
  Johnston et al. (submitted) then went on to show 
  that one consequence of these artificially low
  coronal densities 
  is to (artificially)
  suppress thermal non-equilibrium (TNE) 
  in coronal loops. 
  Thus, under-resolving the TR in
  numerical simulations
  has very significant implications for
  (1)
  the resultant loop dynamics and 
  (2) any comparisons between model predictions and 
  observations.
  \\
  \indent
  Several different approaches have been proposed in order to
  side-step the need for highly resolved numerical grids
  and the commensurate very short time steps that are
  required for numerical stability,
  yet no fully satisfactory solution is available to date.
  For example, \cite{paper:Lionelloetal2009}
  artificially broaden the TR below a fixed cutoff temperature
  ($T_c$), while 
  \cite{paper:Johnstonetal2017a,paper:Johnstonetal2017b}
  dynamically locate the top of an unresolved TR and 
  impose a jump condition that is derived 
  from an integrated form of energy conservation.
  However, 
  when employed in simulations with coarse spatial resolutions,
  the latter approach suffers from overestimating the coronal
  density response to heating
  while the former, in contrast, underestimates the
  density.
  \\
  \indent
  This Letter has presented a new approach, 
  referred to as the TRAC method, that, for the first
  time, has successfully
  removed the influence of under-resolving the TR 
  on the coronal density response to heating
  while retaining remarkable levels of accuracy
  compared 
  with fully resolved models.
  The new method combines the   
  basic ideas from 
  the approaches developed previously  by
  \cite{paper:Lionelloetal2009} and 
  \cite{paper:Johnstonetal2017a,paper:Johnstonetal2017b}.
  \\
  \indent
  We have considered only impulsive heating events where the  
  spatial distribution of the energy deposition 
  is uniform along the loop. 
  However, the 
  TRAC methodology is designed to deal with steep transition 
  regions
  whenever they arise, independent of the nature of the 
  heating. A detailed investigation of different forms of 
  heating
  (e.g.~footpoint heating)
  will be presented in future work.
  \\
  \indent
  TRAC accurately captures the coronal response 
  of properly resolved field-aligned models (e.g.~BC13)
  when employed with spatial resolutions that were
  up to three orders of magnitude coarser. For example,
  with TRAC, grid cell
  widths of order 100 km were sufficient to observe convergence
  to the fully resolved field-aligned model, which required 
  grid cell widths of order 100 m. 
  The peak density errors were less than 3\% 
  and the relaxation of the
  TR resolution culminated in
  computation times that were three orders of magnitude faster
  (e.g.~6 minute run times instead of 3.5 days).
  \\
  \indent
  The advantages of this new approach are multiple. 
  For field-aligned 
  hydrodynamic simulations of the coronal response to heating 
  \citep[see e.g.][for a review]{paper:Reale2014}, 
  the short computation 
  time means that (1) simulations of coronal heating events can 
  be run quickly, permitting extensive surveys of large
  parameter spaces 
  (e.g.~as done by \cite{paper:Fromentetal2018} 
  to study the occurrence of TNE in coronal loops)
  to be completed
  significantly faster
  at a fraction of the computational cost
  and (2) simulations of multiple loop strands 
  (thousands or more) that comprise  an entire active region
  (e.g.~experiments  seeking to reconcile heating models with
  the Hi-C observational data), can be 
  performed with relative ease and high accuracy 
  for the coronal emission.
  However, full numerical resolution is still required
  to deduce the details of the emission in the TR
  at temperatures below 
  the adaptive cutoff temperature.
  \\
  \indent
  In 3D MHD codes, the method can 
  be included without the need for high spatial resolution
  in the TR and a corresponding extended computation time,
  \lq\lq freeing up\rq\rq\ resources to resolve
  better the current
  sheets responsible for the heating.
  Indeed, our 
  results suggest that high levels of 
  accuracy can be obtained with
  grid cell widths of order 100 km, which is achievable 
  in current 3D MHD simulations. 
  %
  %
%
%
  %
  %
  \acknowledgments
  This research has received funding from the 
  European Research Council (ERC) 
  under the European Union's Horizon 2020 research and 
  innovation program (grant agreement No 647214).
  S.J.B. is grateful to the National Science Foundation for 
  supporting this work through CAREER award AGS-1450230.
  C.D.J.
  acknowledges support from the International Space Science 
  Institute (ISSI), Bern, Switzerland to the International Team 
  401 \lq\lq
  Observed Multi-Scale Variability of Coronal Loops as a 
  Probe of Coronal Heating\rq\rq.
  We also thank Dr Z. Miki{\'c} for providing 
  important clarifications on the implementation of the L09 
  method.
  %
  %
%
%
  %
  %
  
  %
  %

\begin{thebibliography}
    \expandafter\ifx\csname natexlab\endcsname\relax\deg  
    \natexlab#1{#1}\fi
    
    \bibitem[Antiochos \& Sturrock(1978)]
    {paper:Antiochos&Sturrock1978}
    {Antiochos}, S.~K. \& {Sturrock}, P.~A.
    \ 1978, \apj, 220, 1137
                   
    \bibitem[Bradshaw \& Cargill(2006)]
    {paper:Bradshaw&Cargill2006}
    {Bradshaw}, S.~J. \& {Cargill}, P.~J.
    \ 2006, \aap, 458, 987
    
    \bibitem[Bradshaw \& Cargill(2010a)]
    {paper:Bradshaw&Cargill2010a}
    {Bradshaw}, S.~J. \& {Cargill}, P.~J.
    \ 2010, \apjl, 710, L39
    
    \bibitem[Bradshaw \& Cargill(2010b)]
    {paper:Bradshaw&Cargill2010b}
    {Bradshaw}, S.~J. \& {Cargill}, P.~J.
    \ 2010, \apj, 717, 163
    
    \bibitem[Bradshaw \& Cargill(2013)]
    {paper:Bradshaw&Cargill2013}
    {Bradshaw}, S.~J. \& {Cargill}, P.~J.
    \ 2013, \apj, 770, 12
    
    \bibitem[Bradshaw \& Klimchuk(2015)]
    {paper:Bradshaw&Klimchuk2015}
    {Bradshaw}, S.~J. \& {Klimchuk}, J.~A.
    \ 2015, \apj, 811, 129
    
     \bibitem[Bradshaw \& Mason(2003)]
    {paper:Bradshaw&Mason2003}
    {Bradshaw}, S.~J. \& {Mason}, H.~E.
    \ 2003, \aap, 407, 1127
    
	\bibitem[Cargill et al.(2012a)]{paper:Cargilletal2012a}
	{Cargill}, P.~J., {Bradshaw}, S.~J. \& {Klimchuk}, J.~A.
	\ 2012, \apj, 752, 161
	
	\bibitem[Cargill et al.(2012b)]{paper:Cargilletal2012b}
	{Cargill}, P.~J., {Bradshaw}, S.~J. \& {Klimchuk}, J.~A.
	\ 2012, \apj, 758, 5

	\bibitem[Cargill et al.(2015)]{paper:Cargilletal2015}
	{Cargill}, P.~J., {Warren}, H.~P. \& {Bradshaw}, S.~J.
	\ 2015, Phil. Trans. Roy. Soc. of Lond. Ser. A, 373, 
	20140260	
	
    \bibitem[Froment et al.(2018)]{paper:Fromentetal2018}
	{Froment}, C., {Auch{\`e}re}, F., {Miki{\'c}}, Z.,
	{Aulanier}, G., {Bocchialini}, K., {Buchlin}, E., 
	{Solomon}, J. \& {Soubri{\'e}}, E. \ 2018, \apj, 855, 52
	
    \bibitem[Johnston et al.(2019)]{paper:Johnstonetal2019}
    {Johnston}, C.~D., {Cargill}, P.~J., {Antolin}, P.,
	{Hood}, A.~W., {De Moortel}, I. 
	\& {Bradshaw}, S.~J.\ 2019, \aap, Submitted
	
    \bibitem[Johnston et al.(2017a)]{paper:Johnstonetal2017a}
    {Johnston}, C.~D., {Hood}, A.~W., {Cargill}, P.~J. \& 
	{De Moortel}, I.\ 2017a, \aap, 597, A81
	
	\bibitem[Johnston et al.(2017b)]{paper:Johnstonetal2017b}
    {Johnston}, C.~D., {Hood}, A.~W., {Cargill}, P.~J. \& 
	{De Moortel}, I.\ 2017b, \aap, 605, A8
	
	\bibitem[Klimchuk et al.(2008)]{paper:Klimchuketal2008}
	{Klimchuk}, J.~A., {Patsourakos}, S. \& 
	{Cargill}, P.~J.\ 2008, \apj, 682, 1351
		
	\bibitem[Linker et al.(2001)]{paper:Linkeretal2001}
    {Linker}, J.~A., {Lionello}, R., {Miki{\'c}}, Z. \&
    {Amari}, T.\ 2001, \jgr, 106, 25165
	
    \bibitem[Lionello et al.(2009)]{paper:Lionelloetal2009}
    {Lionello}, R., {Linker}, J.~A. \& {Miki{\'c}}, Z.
    \ 2009, \apj, 690, 902
	
	\bibitem[Miki{\'c} et al.(2013)]{paper:Mikicetal2013}
    {Miki{\'c}}, Z., {Lionello}, R., {Mok}, Y., {Linker}, J.~A.
    \& {Winebarger}, A.~R.\ 2013, \apj, 773, 94
    
    \bibitem[Reale(2014)]{paper:Reale2014}
    {Reale}, F., \ 2014, Liv. Rev. Sol. Phys., 11 
    
    \bibitem[Reale(2016)]{paper:Reale2016}
    {Reale}, F., \ 2016, \apjl, 826, L20
    
  \end{thebibliography}
\end{document}